\documentclass[12pt,onecolumn,a4paper, draftcls]{IEEEtran}
\usepackage{graphicx}
\usepackage{epstopdf}

\usepackage[utf8]{inputenc}
\usepackage[T1]{fontenc}
\usepackage[english]{babel}
\usepackage{amssymb}
\usepackage[T1]{fontenc}
\usepackage{hyperref}
\usepackage[english]{babel}
\usepackage{url}
\usepackage{multirow}
\usepackage{float}
\usepackage{cite}
\usepackage{amssymb,amsmath}
\usepackage[]{algorithm}
\usepackage[]{algorithmic}
\usepackage{multicol,lipsum}
\usepackage{mathtools,esdiff}
\usepackage{booktabs}

\usepackage{soul}  

\usepackage{bm}
\usepackage{subcaption}
\usepackage{tikz}
\definecolor{gold}{rgb}{0.85,.66,0}
\newcommand{\colb}{\textcolor{black}}

\begin{document}
	
\title{Improving Spectral Efficiency via Pilot Assignment and Subarray Selection Under Realistic XL-MIMO Channels}
\author{{Gabriel Avanzi Ubiali}, {Taufik Abrão}, José Carlos Marinello\\
\thanks{This work was partly supported by the National Council for Scientific and Technological Development (CNPq) of Brazil under Grants 310681/2019-7 and  163568/2021-9 (scholarship), and in part by State University of Londrina (UEL), PR, Brazil.}
\thanks{G. A. Ubiali and T.  Abrão are with the Electrical Engineering Department, State University of Londrina, PR, Brazil.  Rod. Celso Garcia Cid - PR445,  s/n, Campus Universitário, Po.Box 10.011.  CEP: 86057-970. E-mail: \url{g.avanziubiali@gmail.com}; \quad \url{taufik@uel.br}}
\thanks{J. C. Marinello is with Electrical Engineering Department, Federal University of Technology PR, Cornélio Procópio, PR, Brazil. CEP: 86300-000.  \url{jcmarinello@utfpr.edu.br}.}
}
	
\maketitle

\begin{abstract}
The main requirements for 5G and beyond connectivity include a uniform high quality of service, which can be attained in crowded scenarios by extra-large MIMO (XL-MIMO) systems. Another requirement is to support \colb{an increasing number of} connected users in (over)crowded machine-type communication (mMTC). In such scenarios, pilot assignment (PA) becomes paramount to reduce pilot contamination and consequently improve spectral efficiency (SE).  We propose a novel quasi-optimal low-complexity iterative pilot assignment strategy for XL-MIMO systems, based on a genetic algorithm (GA). The proposed GA-based PA procedure turns the quality of service more uniform, taking into account the {normalized mean-square error} (NMSE) of channel estimation from each candidate of the population. The simulations reveal that the proposed iterative procedure minimizes the channel estimation NMSE averaged over the UEs.
The second 
procedure is the subarray (SA) selection. In XL-MIMO systems, {commonly} a UE is close to a SA antenna subset such that a sufficient data rate can be achieved if only a specific SA serves that UE. Thus, a SA selection procedure is investigated to make the system scalable by defining the maximum number of UEs each SA can help.  Hence, the SA clusters are formatted based on the PA decision. Furthermore, we introduce an appropriate channel model for XL-MIMO, which considers a deterministic LoS component with a distance-dependent probability of existence combined with a stochastic spatially correlated Rayleigh NLoS fading component. The developed simulations and analyses rely on this suitable channel model under realistic assumptions of pilot contamination and correlated channels.
\end{abstract}
	
\begin{IEEEkeywords}
Extra-large-scale antenna array, non-stationary channels, channel model, spectral efficiency, pilot assignment, subarray selection
\end{IEEEkeywords}

\section{Introduction}
\label{sec:Intro}

One of the main enabling technologies for achieving high data rates in wireless networks is {\it Massive multiple-input multiple-output} (mMIMO). The massive number of channel observations, at the AP antennas, increases the array gain and the ability of using the spatial domain to discriminate the desired signal from the interfering signals, in the {\it uplink} (UL), and to transmit the signals very directively, in the {\it downlink} (DL). Indeed, the UEs that are close to one of the APs, {\it i.e.}, located at a {\it cell-center area}, will experience high data rates; however, the UEs that are relatively distant from any AP, {\it i.e.}, located at a {\it cell-edge area}, may have bad quality of service (QoS) due to low {\it signal-to-noise-ratio} (SNR) and to the inter-cell interference, which comes from the UEs in the neighboring cells, due to the reuse of pilot sequences and to the inability of an AP to cancel the interference from the other cells' UEs, since it only estimates the channel of the UEs it is serving.

Thus, despite the current cellular networks can achieve high peak data rates, the QoS could result in unreliable due to significant data rate variations within the cell. However, wireless access is supposed to be ubiquitous since payments, navigation, and entertainment rely on wireless connectivity. It is more important for future mobile networks to provide more uniform high rates over the coverage area than to increase the peak rates. More uniform high rates can be achieved under large antenna arrays dimensionality, the so-called {\it extra-large} MIMO (XL-MIMO) networks, which can serve relatively large areas such as stadiums and shopping malls densely populated. XL-MIMO networks can turn the QoS more uniform across the entire cell area by achieving moderate to high macro-diversity and reducing the average distance from a UE to the closest AP antenna.

An XL-MIMO antenna array is split into {\it subarray} (SA) arrangements. As each UE is physically close to a subset of SAs only, the configuration where all the SAs serve all the UEs is impractical and unnecessary. Limiting the set of SAs that serve each UE can \colb{negatively affect} the QoS, but if the SA selection is made judiciously, the \colb{system can still provide suitable SE, while also reducing the computational complexity and the} amount of signaling and control data that needs to be transferred from the SAs to the CPU and vice-versa. Furthermore, SA selection procedure is essential in XL-MIMO systems to guarantee scalability, {\it i.e.}, to assure that the complexity will not go to infinity as the number of UEs grows unbounded.

Knowledge of the UE--SA channel responses is paramount to coherent processing. The most efficient way to estimate the channels is to consider a TDD scheme where we only need to transmit pilots in the uplink. Depending on the users' mobility, the size of the normalized coherence block may be smaller than the number of UEs. The mutually orthogonal pilot sequences must be reused among the UEs in this case. Most of XL-MIMO papers consider perfect channel estimation \cite{AMIRI2019,Nishimura2020,Marinello2020,UL_XLMIMO_2020,ExpPropDetector_XL-MIMO,AMIRI2022}, which is unrealistic. During the UL pilot transmission, the estimation error is generated by two sources: the background noise and the signals coming from the pilot-sharing UEs. The last one results in so-called coherent interference. In this paper, we develop our analysis assuming imperfect channel estimations to obtain realistic numerical simulation results.

The XL-MIMO channel can be described as the composition of a deterministic {\it line-of-sight} (LoS) path and a stochastic {\it non}-LoS (NLoS) component. 
Due to the low probability of LoS occurrence, generally, the LoS component modeling is not  distinguished between suburban and urban macro scenarios; indeed, it is selectable for the urban micro scenario only \cite{3GPP.TR.25.996}. Usually, research papers in mMIMO only consider the NLoS channel component. On the other hand, papers in CF-mMIMO usually consider either LoS or NLoS component. However, due to the uncertainties in the propagation environment in urban areas, it is impossible to accurately determine the presence of a LoS link\cite{LOS_NLOS_2022}, and a probabilistic LoS model is necessary. Furthermore, the LoS component contributes considerably to the total channel power\cite{[2022]Cell-Free_Massive_MIMO_A_Survey}, since it is associated to a shorter propagation distance and to a slower signal attenuation with the distance. Particularly, in XL-MIMO systems, the distance between a UE and the BS antenna array may be small relative to the XL-array length, so the LoS probability should not be neglected. In this paper, we introduce a realistic channel model that incorporates a deterministic LoS component, with a distance-dependent probability of existence, and a spatially correlated NLoS component. Shadow fading correlation between two different links is also considered.


Depending on the mobility, the size of the coherence block may be smaller than the number of UEs. In this case, the mutually orthogonal pilots must be reused among the UEs and can be appropriately assigned to them to limit the pilot contamination\cite{CFbook}. The most simple and naive approach is to give pilots randomly. However, it cannot prevent closely located UEs from being assigned the same pilot, leading to high interference and system performance degradation. On the other extreme, the exhaustive search, which tries all possible assignments and chooses the best one, leads to a computational complexity that presents a combinatorial growth with the number of active UEs. Thus, only suboptimal methods are feasible in practice. One approach is to optimize a utility function concerning one UE at a time, considering each UE once or iterating until convergence \cite{CFbook}. Several suboptimal algorithms consider a golden rule that closely located UEs should not be assigned the same pilot.

Works on cellular mMIMO usually consider that two UEs within the same cell are assigned mutually orthogonal pilot sequences. If the pilot reuse factor is one, the performance of cell-edge UEs may be degraded by pilot-sharing UEs of adjacent cells. In this scenario, the pilot contamination problem can be reasonably relieved greedily by simply assigning the best pilots, {\it i.e.}, the ones that result in the most negligible pilot contamination to the cell-edge UEs\cite{zhu2015smart}. \colb{In \cite{Omid2021}, the problem of pilot assignment in multi-cell mMIMO systems is tackled by using a {\it deep reinforcement learning} (DRL) scheme. The distance and the angle-of-arrival (AoA) information of the users are deployed to define the cost function, representing the pilot contamination effects. The DRL algorithm is applied to ﬁnd an effective pilot assignment strategy by tracking the changes in the environment, learning the near-optimal pilot assignment, and achieving comparable performance to that of the optimum (genie) PA performed by exhaustive search.}

\colb{However, due to the increased density of UEs in typical applications of XL-MIMO systems and the scarcity of pilots, crucial tasks, such as pilot assignment, network access, and user scheduling, become challenging \cite{Marinello2022}. In our} work, we consider a crowded XL-MIMO scenario where the number of UEs in the coverage area is more significant than the number of orthogonal pilot sequences\colb{, representing a scenario with pilot scarcity}. A greedy algorithm would assign the orthogonal pilots randomly to some of the UEs, and then assign pilots to the remaining UEs to minimize the pilot contamination, each UE at a time.
Works in \cite{Nishimura2020, Marinello2022, Cui2022, Nishimura2022} address the problem of random access and pilot assignment in XL-MIMO, considering uncorrelated Rayleigh fading and no subarray selection. \colb{Specifically, a joint random access  and scheduling 
protocol was proposed in \cite{Marinello2022}, taking advantage of the different visibility regions (VRs) that arise  in the UE and XL-MIMO link, besides seeking UEs with non-overlapping VRs to be scheduled in the same payload data pilot resource. The difference to our work is that we focus on the pilot assignment problem and develop an efficient PA algorithm for a probabilistic LoS/NLoS XL-MIMO channel model based on a heuristic evolutionary genetic algorithm (GA) approach} and build up the analysis considering the employment of SA selection to make the receive combining scalable.

The contribution of this work is threefold.
\begin{itemize}
\item[{\it i})] We introduce a realistic XL-MIMO spatially {\it correlated channel model}, since uncorrelated channels appear under extremely strict physical requirements. Since the XL-MIMO antennas are not co-located, sometimes the UEs will be very close to some SAs, such that the channel modeling must be a composition of a sum of NLoS links and an LoS link, with a distance-dependent probability of occurrence. To the best of our knowledge, all the works on XL-MIMO have considered only one of these two components. 
\item[{\it ii})] We propose a GA-based PA strategy for crowded XL-MIMO systems, mitigating pilot contamination by approaching the optimal PA solution w.r.t. minimizing the average {\it normalized mean-square error} (NMSE) of the channel estimates among all the active UEs.
\item[{\it iii})] We provide extensive numerical results corroborating the effectiveness and efficiency of the proposed SA selection and PA methodology for XL-MIMO systems, exploring crowded scenarios by considering both the LoS and NLoS channel components. All the analysis and obtained results of this paper are based on the realistic assumption of imperfect channel estimation, the employment of a scalable receive combiner, and a proper SA selection, which is also essential to make the XL subarrays scalable.
\end{itemize}

The remainder of this paper is organized as follows. Section \ref{sec:ChannelModel} describes the proposed channel model for XL-MIMO. Section \ref{sec:SystemModel} deals with the XL-MIMO system model, describing the signal model for the UL pilot and data transmissions and the adopted channel estimator and receive combiner. Section \ref{sec:PA_SAS} presents the considered strategies for PA in XL-MIMO and the adopted SA selection algorithm. Finally, numerical results corroborating our findings are provided in Section \ref{sec:Results}, and final remarks and conclusions are presented in Section \ref{sec:Conclusions}.


\section{XL-MIMO Channel Model}
\label{sec:ChannelModel}

\begin{figure}[htbp!]
\centering
\subfloat[\colb{3-D view of the XL-MIMO system}]{\label{fig:XLMIMO_3D}{\includegraphics[trim={0mm 0mm 1mm 0mm},clip,width=.5\linewidth]{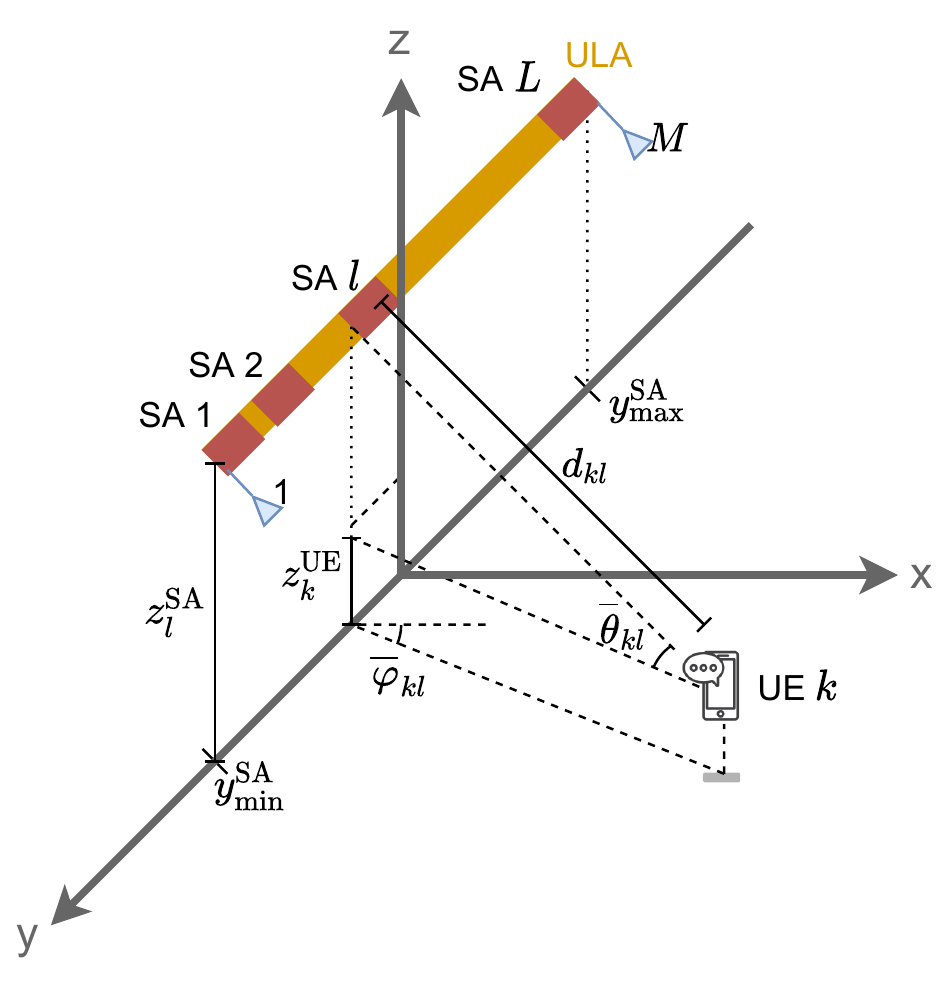}}}\hfill
\subfloat[\colb{The $l$-th SA and $k$-th UE projected in plane $z=0$}]{\label{fig:XLMIMO_SA_2D}{\includegraphics[trim={0mm 0mm 1mm 0mm},clip,width=.5\linewidth]{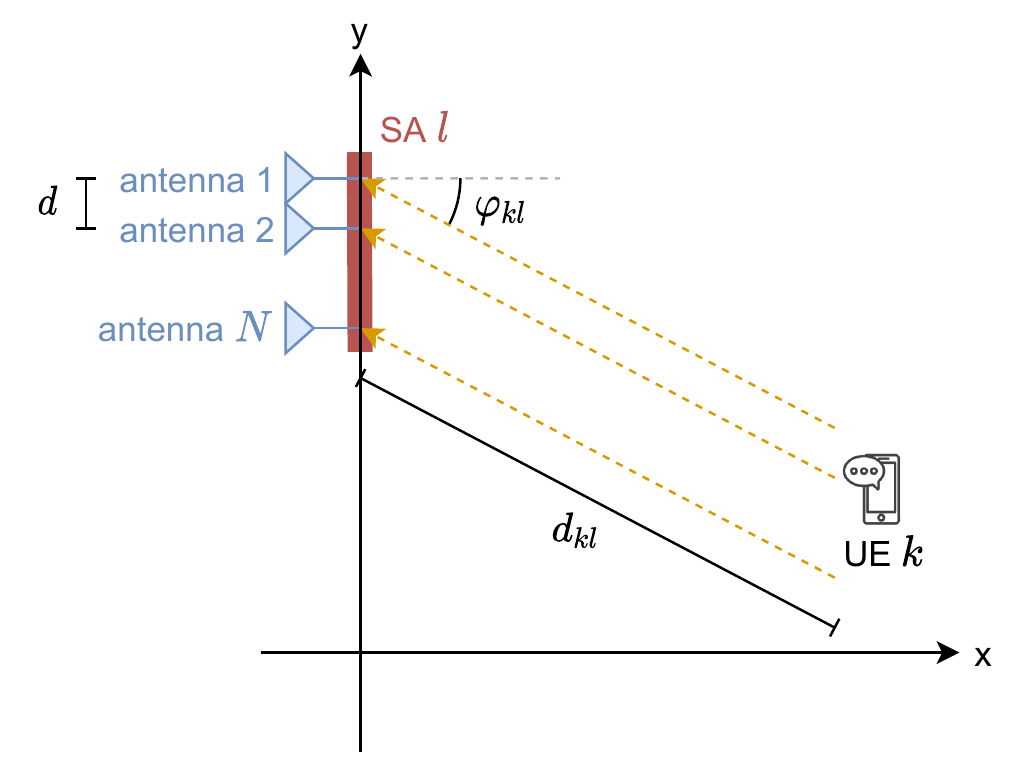}}}\hfill
\caption{\colb{3-D XL-MIMO representation, showing a) the ULA and the azimuth and elevation angles corresponding to the LoS path between UE $k$ and SA $l$ and b) the antenna spacing $d$ (inside SA $l$) and the distance between UE $k$ and SA $l$.}}
\label{fig:angles_and_distances}
\end{figure}

We consider an XL-MIMO antenna array with a $M$-antenna ULA, divided into $L$ SAs containing $N$ antennas each, such that $M=LN$, serving $K$ UEs. The antennas in each SA are separated by a distance $d$. If $N$ is small, then each SA will experience spatial stationarity, {\it i.e.}, all its $N$ antennas will experience the same condition of existence or absence of the LoS path component. Each SA is connected to a central processing unit (CPU), which is responsible for the SA cooperation. The height of the $l$th SA and the $k$th UE are denoted by $z_l^\text{SA}$ and $z_k^\text{UE}$, respectively, and $d_{kl}$ denotes the distance from the $k$th UE to the first antenna of the $l$th SA. Fig. \ref{fig:angles_and_distances} illustrates the XL-MIMO antenna array (yellow bar) highlighting the azimuth and elevation angles associated to the LoS path between a given UE $k$ and a given SA $l$, represented by the red bar.

The associated channel model with LoS and the NLoS components can be expressed as:
\begin{equation}
\textbf{h}_{kl} = \alpha_{kl}\textbf{h}_{kl}^\text{LoS} + \textbf{h}_{kl}^\text{NLoS},
\label{h_kl}
\end{equation}
where $\textbf{h}_{kl}^\text{LoS}, \textbf{h}_{kl}^\text{NLoS},\textbf{h}_{kl}\in\mathbb{C}^N$ denote, respectively, the LoS, the NLoS and the probabilistic LoS/NLoS channel vectors between UE $k$ and SA $l$, while $\alpha_{kl}$ is a Bernoulli random variable indicating the presence ($\alpha_{kl}=1$) or absence ($\alpha_{kl}=0$) of a LoS component between the $k$th UE and the $l$th SA.

The LoS channel vector between the $k$th UE and the $l$th SA is given by\cite{LOS_NLOS_2022}
\begin{equation}
\textbf{h}_{kl}^\text{LoS} = \sqrt{G_kG_l}\frac{z_k^\text{UE}z_l^\text{SA}}{4\pi d_{kl}}\sqrt{\mathcal{X}_{kl}^\text{LoS}}e^{-j2\pi \frac{d_{kl}}{\lambda}}\textbf{a}(\overline{\varphi}_{kl},\overline{\theta}_{kl}),
\label{h_LoS_complete}
\end{equation}
where $G_k$ and $G_l$ denote the antenna gain at the $k$th UE and the $l$th SA, respectively; $\lambda=c/f$ is the wavelength, being $c$ the speed of light and $f$ the carrier frequency; $\mathcal{X}_{kl}^\text{LoS}$ is the shadow fading of the LoS component; and $\textbf{a}(\overline{\varphi}_{kl},\overline{\theta}_{kl})$ is the array response vector of the ULA, regarding to the LoS ray. The azimuth and elevation angles of the LoS path between the $k$th UE and the $l$th SA are denoted by $\overline{\varphi}_{kl}$ and $\overline{\theta}_{kl}$, respectively.
\colb{The {\it array response vector}, as a function of the LoS azimuth and elevation angles between UE $k$ and SA $l$, respectively $\overline{\varphi}_{kl}$ and $\overline{\theta}_{kl}$, is given by}
\begin{equation}
\colb{
\textbf{a}(\overline{\varphi}_{kl},\overline{\theta}_{kl}) = \left[1,e^{-j2\pi\frac{d}{\lambda}\frac{\text{sin}(\overline{\varphi}_{kl})}{{\text{cos}(\overline{\theta}_{kl})}}},\dots,e^{-j2(N-1)\pi\frac{d}{\lambda}\frac{\text{sin}(\overline{\varphi}_{kl})}{{\text{cos}(\overline{\theta}_{kl})}}}
\right]^\text{T},
}
\label{a}
\end{equation}
%
\colb{where $d$ is the antenna spacing inside the $l$-th SA, Fig. \ref{fig:XLMIMO_SA_2D}. Indeed, the array response vector $\textbf{a}(\cdot,\cdot)$ in eq. \eqref{a}} accounts for the phase differences among the signal that arrives in different antennas of the same {\it uniform linear SA} (ULSA). These variations appear when the azimuth angle is not zero, {\it i.e.}, when the wavefront is not parallel to the ULA.

\vspace{2mm}
\noindent\textbf{\textit{Near-field vs. Far-field}}.
An important observation on XL-MIMO arrays is that the wavefront propagation \colb{could} result in spherical, since some users may be in the near-field distance w.r.t. the entire XL-MIMO array, {{\it i.e.}, $d_{kl}\ll d_{\rm Rayl}$, being $d_{\rm Rayl}$ the Rayleigh distance \cite{Kraus2002Antennas}.} \colb{Also, such conditions will result} in different propagation distances for different SAs. On the other hand, since the SA aperture is small compared to the \colb{UE-SA propagation distances}, \colb{one can assume that} all users are in the \colb{far-field propagation} from each SA perspective, such that we can assume that the wavefront \colb{propagation} is {\it planar} to the SA point of view. The implication is that the difference between the propagation distance for two antennas belonging to the same SA is negligible, such that the corresponding large-scale fading (LSF) coefficients are nearly the same \colb{for all antenna elements of the same SA}. However, even this negligible difference is sufficient to produce phase differences among these antennas. Such phase differences are accounted by the array response vector in \eqref{a}.

The LSF channel gain for the LoS component in each XL-MIMO ULSA is determined by the normalized trace as
\begin{equation}
\beta_{kl}^\text{LoS} = \frac{1}{N}||\textbf{h}_{kl}^\text{LoS}||^2 = \underbrace{\frac{G_kG_l(z_k^\text{UE}z_l^\text{SA})^2}{(4\pi)^2}}_{\beta_0}\frac{\mathcal{X}_{kl}^\text{LoS}}{d_{kl}^2}=\frac{\beta_0}{d_{kl}^2}\mathcal{X}_{kl}^\text{LoS},
\label{beta_LoS}
\end{equation}
where $\beta_0$ is the average channel gain at the reference distance of 1 m, and $\beta_0/d_{kl}^2$ is the path-loss of the LoS component. Replacing \eqref{beta_LoS} in \eqref{h_LoS_complete} gives
\begin{equation}
\textbf{h}_{kl}^\text{LoS} = \sqrt{\beta_{kl}^\text{LoS}}e^{-j2\pi \frac{d_{kl}}{\lambda}}\textbf{a}(\overline{\varphi}_{kl},\overline{\theta}_{kl}).
\label{h_LoS}
\end{equation}

On the other hand, the NLoS component describes a practical {\it spatially correlated multi path} environment. The NLoS channel vector can be obtained as the superposition of $B$ physical signal paths, each \colb{one} reaching the ULSA as a plane wave from particular azimuth and elevation angles $\varphi_{kl}^{(b)}$ and $\theta_{kl}^{(b)}$ \cite{[2020]Toward_Massive_MIMO_2.0_Understanding_Spatial_Correlation_Interference_Suppression_and_Pilot_Contamination}:
\begin{equation}
\textbf{h}_{kl}^\text{NLoS} = \sum_{b=1}^B g_{kl}^{(b)}\textbf{a}(\varphi_{kl}^{(b)},\theta_{kl}^{(b)}),
\label{h_NLoS}
\end{equation}
where $g_{kl}^{(b)}$ is an i.i.d. complex random variable with zero mean and variance $\mathbb{E}\{|g_{kl}^{(b)}|^2\}$, accounting for the gain and phase rotation of the $b$th physical NLoS path, while $\varphi_{kl}^{(b)}$ and $\theta_{kl}^{(b)}$ are, respectively, the SA azimuth and elevation angles formed by the $b$th NLoS path between UE $k$ and SA $l$.

The NLoS LSF coefficient can be calculated as
\begin{equation}
\beta_{kl}^\text{NLoS} = \frac{1}{N}\mathbb{E}\{||\textbf{h}_{kl}^\text{NLoS}||^2\} = \sum_{b=1}^B\mathbb{E}\{|g_{kl}^{(b)}|^2\},
\label{beta_NLoS_def}
\end{equation}
and is given by
\begin{equation}
\beta_{kl}^\text{NLoS}=\frac{\beta_0}{d_{kl}^\gamma}\mathcal{X}_{kl}^\text{NLoS},
\label{beta_NLoS}
\end{equation}
where $\gamma$ is the path-loss exponent of the NLoS component, and $\mathcal{X}_{kl}^\text{NLoS}$ is the shadow fading of the NLoS component. Considering the law of large numbers, the sum of the $B$ random NLoS paths approaches a normal distribution\cite{[2020]Toward_Massive_MIMO_2.0_Understanding_Spatial_Correlation_Interference_Suppression_and_Pilot_Contamination}:
\begin{equation}
\textbf{h}_{kl}^\text{NLoS} \sim \mathcal{N}_\mathbb{C}(\mathbf{0},\textbf{R}_{kl}),
\label{h_NLoS_distribution}
\end{equation}
with the positive semi-definite {\it spatial correlation matrix} $\textbf{R}_{kl}$ composed by the elements:
\begin{equation}
[\textbf{R}_{kl}]_{m,n} = \beta_{kl}^\text{NLoS}\int\int e^{j2(m-n)\pi\frac{d}{\lambda}\frac{\text{sin}(\varphi)}{\text{cos}(\theta)}}f_{kl}(\varphi,\theta)d\varphi d\theta,
\label{R}
\end{equation}
where $f_{kl}(\varphi,\theta)$ is the joint probability density function (jPDF) of $\varphi$ and $\theta$, which depends on the azimuth and elevation angles of the LoS link ($\varphi_{kl}$ and $\theta_{kl}$) and on the angular standard deviations ($\sigma_\varphi\geq0$ and $\sigma_\theta\geq0$). The integration interval for $\varphi$ and $\theta$ is $[-180^{\circ},180^{\circ}]$ and $(-90^{\circ},90^{\circ})$, respectively. As in \cite{CFbook}, we also consider the joint Gaussian distribution:
\begin{equation}
f_{kl}(\varphi,\theta) = \frac{1}{2\pi\sigma_\varphi\sigma_\theta}
\exp\left\{-\frac{(\varphi-\overline{\varphi}_{kl})^2}{2\sigma_\varphi^2} -\frac{(\theta-\overline{\theta}_{kl})^2}{2\sigma_\theta^2}\right\}.
\label{PDF_angles}
\end{equation}

The proof of \eqref{h_NLoS_distribution} and \eqref{R} can be found in Appendix \ref{app1}. 
From \eqref{R}, notice that $\text{tr}(\textbf{R}_{kl})=N\beta_{kl}^\text{NLoS}$, which matches with the definition in \eqref{beta_NLoS_def}.

Separating the large-scale and the small-scale fading, the NLoS channel vector can be described as
\begin{equation}\label{eq:h_NLoS_eq}
\textbf{h}_{kl}^\text{NLoS} = \sqrt{\beta_{kl}^\text{NLoS}} \cdot \ddot{\textbf{h}}_{kl},
\end{equation}
where $\ddot{\textbf{h}}_{kl}$ represents the {\it spatially correlated small-scale} Rayleigh {\it fading} following a complex normal distribution\cite{massivemimobook,CFbook,[2020]Toward_Massive_MIMO_2.0_Understanding_Spatial_Correlation_Interference_Suppression_and_Pilot_Contamination}. Note that the double-integral in \eqref{R} computes $\mathbb{E}\left\{e^{j2(m-n)\pi\frac{d}{\lambda}\frac{\text{sin}(\varphi)}{\text{cos}(\theta)}}\right\}$ w.r.t. randomly located multipath components distributed according to the {joint} PDF in \eqref{PDF_angles}.

In practical XL-MIMO panel scenarios, as the antennas of different SAs are sufficiently distant, we assume the channel vectors of other SAs are independently distributed, which means that $\mathbb{E}\{\ddot{\textbf{h}}_{kl}\ddot{\textbf{h}}_{kj}^\text{H}\}= \mathbf{0}_{N\times N}$ for $l\neq j$, where $(.)^\text{H}$ denotes the Hermitian operator.

%

\vspace{2mm}
\noindent\textbf{\textit{Correlation between $F_{kl}$ and $F_{ij}$ (Shadowing)}}.
The {\it shadow fading} can be modeled as a log-normal r.v., 
in dB scale as
$F_{kl}=10~\text{log}_{10}(\mathcal{X}_{kl})$, following the distribution  \cite{[2015]Effects_of_Correlated_Shadowing_Modeling_on_Performance_Evaluation_of_Wireless_Sensor_Networks}
\begin{equation}\label{eq:F_kl}
F_{kl} \sim \mathcal{N}\left[0,\sigma_\text{SF}^2(1-e^{-d_{kl}/\delta})^2\right].
\end{equation}

Note that the variance of the r.v. $F_{kl}$ in \eqref{eq:F_kl} increases with the distance and converges to $\sigma_\text{SF}^2$ when $d_{kl}>>\delta$, where $\delta$ denotes the decorrelation distance. The shadow fading is typically modeled as independent links, although in reality proximate links present spatially correlated shadowing, since the signal is impacted by almost the same set of objects in the propagation environment. Ignoring such correlations results in over-estimating the diversity in XL-MIMO systems and may produce significant differences in network performance\cite{[2015]Effects_of_Correlated_Shadowing_Modeling_on_Performance_Evaluation_of_Wireless_Sensor_Networks}. The cross-correlation between $F_{kl}$ and $F_{ij}$ is given by: 
\begin{align}
\mathbb{E}\{F_{kl}F_{ij}\}
=\frac{\sigma_\text{SF}^2}{2}\frac{(1-e^{-d_{kl}/\delta})(1-e^{-d_{ij}/\delta})}{\sqrt{(1+e^{-d_{kl}/\delta})(1+e^{-d_{ij}/\delta})}}(e^{-d_{kj}/\delta}+e^{-d_{il}/\delta}+e^{-D_{ki}^\text{UE}/\delta}+e^{-D_{lj}^\text{SA}/\delta}).
\label{cross_corr}
\end{align}
%
where $d_{kl}$ is the distance between the $k$th UE and the $l$th SA, $D_{ki}^\text{UE}$ is the distance between UE $k$ and UE $i$, and $D_{lj}^\text{SA}$ is the distance between SA $l$ and SA $j$ \cite{[2014]Efficient_modeling_of_correlated_shadow_fading_in_dense_wireless_multi-hop_networks,[2015]Effects_of_Correlated_Shadowing_Modeling_on_Performance_Evaluation_of_Wireless_Sensor_Networks}. Note that the closer the $k$th UE is from the $i$th UE, the more correlated their shadow fading, $F_{kl}$ and $F_{ij}$, will be. Also, the closer the $l$th SA is from the $j$th SA, the more correlated $F_{kl}$ and $F_{ij}$ will be.

Finally, from \eqref{h_kl} and \eqref{h_NLoS_distribution}, the channel vector follows a {\it complex normal distribution} with mean $\overline{\textbf{h}}_{kl}=\alpha_{kl}\textbf{h}_{kl}^\text{LoS}$ and {\it covariance matrix}
\begin{align}
\mathbb{C}\text{ov}(\textbf{h}_{kl})
=&~ 
\mathbb{E}\{\textbf{h}_{kl}\textbf{h}_{kl}^\text{H}\}-\overline{\textbf{h}}_{kl}\overline{\textbf{h}}_{kl}^\text{H}
\notag\\ 
=&~  \mathbb{E}\{(\overline{\textbf{h}}_{kl}+\textbf{h}_{kl}^\text{NLoS})(\overline{\textbf{h}}_{kl}+\textbf{h}_{kl}^\text{NLoS})^\text{H}\}-\overline{\textbf{h}}_{kl}\overline{\textbf{h}}_{kl}^\text{H}
\notag\\ 
=&~
\mathbb{E}\{\textbf{h}_{kl}^\text{NLoS}(\textbf{h}_{kl}^\text{NLoS})^\text{H}\}\notag\\ 
=&~ \textbf{R}_{kl},
\label{R_kl}
\end{align}
%
%
such that
\begin{equation}
\textbf{h}_{kl} \sim \mathcal{N}_\mathbb{C}(\overline{\textbf{h}}_{kl},\textbf{R}_{kl}).
\label{h_distribution}
\end{equation}
Notice that the correlation matrix of $\textbf{h}_{kl}^\text{NLoS}$ is also $\textbf{R}_{kl}$, since $\textbf{h}_{kl}^\text{NLoS}$ is a zero mean vector. Indeed, as demonstrated in \eqref{R_kl}, the covariance matrix of $\textbf{h}_{kl}$ is also $\textbf{R}_{kl}$; however, its correlation matrix is $\textbf{R}_{kl}+\overline{\textbf{h}}_{kl}\overline{\textbf{h}}_{kl}^\text{H}$, since it is a non-zero mean vector.

Both $\overline{\textbf{h}}_{kl}$ and $\textbf{R}_{kl}$ describe the macroscopic propagation effects, {\it i.e.}, path-loss and shadowing, whereas the distribution in \eqref{h_distribution} describes the small-scale fading. If there is a LoS path between UE $k$ and SA $l$, then we say that SA $l$ is inside the visibility region (VR) of UE $k$. Otherwise, the received power is much weaker as the signal is coming only from NLoS paths, and then we say that SA $l$ is not in the VR of UE $k$.

Finally, the probabilistic LoS/NLoS LSF channel gain is defined as
\begin{equation}
\beta_{kl} = \frac{1}{N}\mathbb{E}\{||\textbf{h}_{kl}||^2\},
\label{beta_def}
\end{equation}
and, by replacing \eqref{h_kl} into \eqref{beta_def}, and recalling that $||\textbf{h}_{kl}^\text{LoS}||^2=N\beta_{kl}^\text{LoS}$ and $\mathbb{E}\{||\textbf{h}_{kl}^\text{NLoS}||^2\}=\text{tr}(\textbf{R}_{kl})=N\beta_{kl}^\text{NLoS}$, we come to
\begin{equation}
\beta_{kl} = \alpha_{kl}\beta_{kl}^\text{LoS} + \beta_{kl}^\text{NLoS}.
\label{beta}
\end{equation}

\begin{figure}[htbp!]
\centering
\includegraphics[trim={0mm 0mm 0mm 0mm} ,clip,width=.66\linewidth]{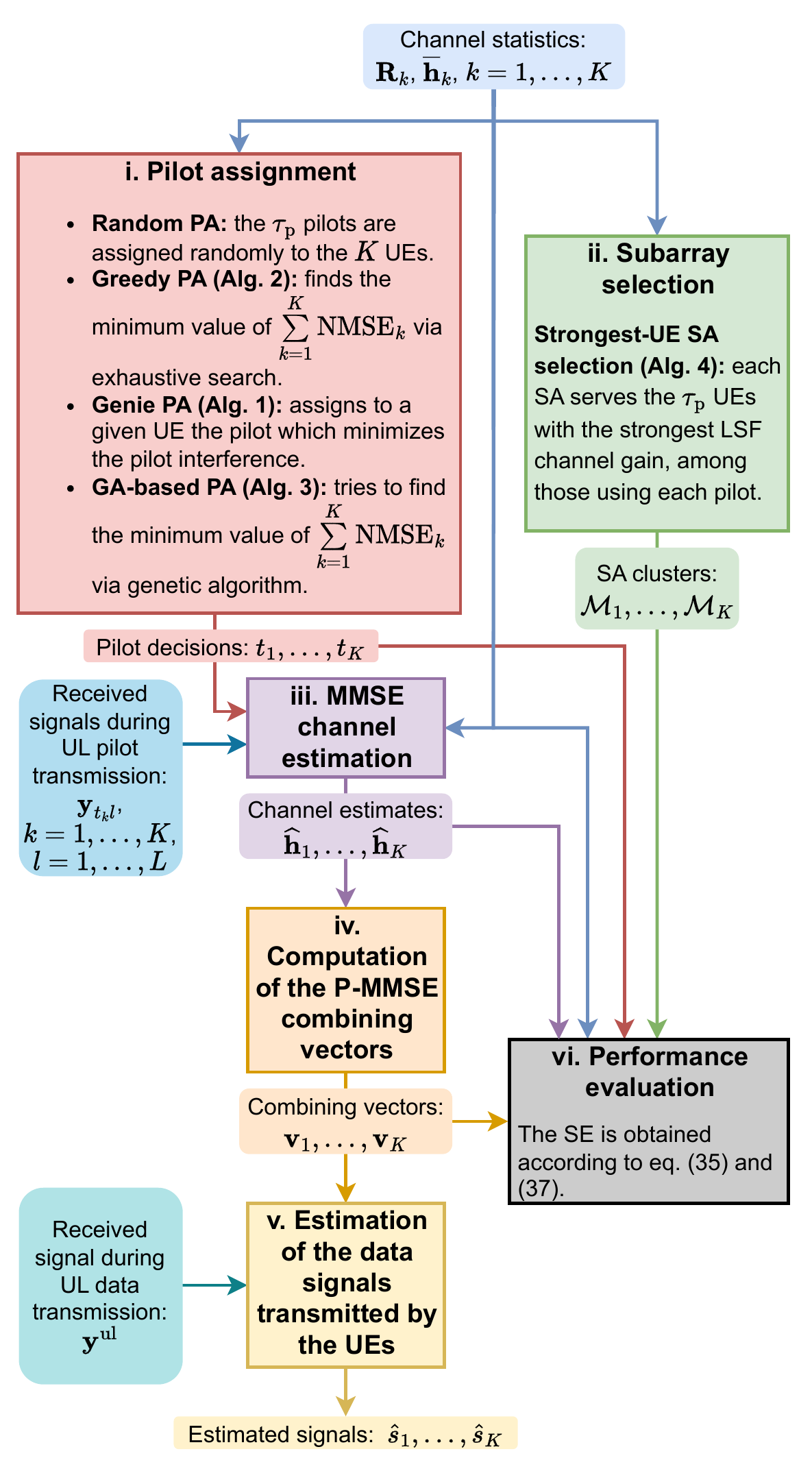}
\vspace{-4mm}
\caption{System operation phases: {\it i}) \colb{pilot assignment (four methods were compared)}; {\it ii}) subarray selection; {\it iii}) channel estimation; {\it iv}) \colb{computation of the combining vectors}; \colb{{\it v}) signal estimation; {\it vi}) SE calculation.}} 
\label{fig:fluxogram}
\end{figure}

In the following sections, we describe the XL-MIMO system model and the procedures for both PA and SA selection. An overview of the operations involved in such processes is shown in  Fig. \ref{fig:fluxogram}\colb{: first, the pilots are assigned to the $K$ UEs based on the channel statistics (channel average and covariance matrices). Second, the SA selection is performed, also based on the channel statistics. Third, the instantaneous channels are estimated in each coherence block, based on the pilot decisions and the received signals during the UL pilot transmission. Fourth, the combining vectors are computed, based on the channel statistics and the pilot decisions (which are necessary to obtain the channel estimation covariance matrices) and on the channel estimates. Fifth, the signals transmitted by the UEs are estimated, based on the combining vectors and on the received signal at the ULA during the UL data transmission. Finally, when evaluating the system performance, we can compute the SINR  based on eq. \eqref{SINR_k_centralized}, based on the channel estimates, channel statistics, pilot decisions, SA clusters and combining vectors.}

\vspace{-5mm}

\newpage
\section{XL-MIMO System Model}
\label{sec:SystemModel}

\subsection{UL Pilot Transmission}
\label{subsec:PilotTransmission}

Pilots are $\tau_\text{p}$-dimensional vectors, and we can only find at most $\tau_\text{p}$ mutually orthogonal sequences. The set $\boldsymbol{\phi}_1,\dots,\boldsymbol{\phi}_{\tau_\text{p}}\in\mathbb{C}^{\tau_\text{p}}$, containing $\tau_\text{p}$ orthogonal pilot sequences, is defined such that
\begin{equation}
\boldsymbol{\phi}_i^\text{H}\boldsymbol{\phi}_j = 
\begin{cases}
\tau_\text{p}, \quad & \text{if}\ i=j,\\
0, \quad & \text{if}\ i\neq j.
\end{cases}
\label{pilot_sequences_orthogonality}
\end{equation}

The channels need to be estimated once per coherence block. Therefore, $\tau_\text{p}$ samples must be reserved for uplink pilot signaling in each coherence block, but the finite length of the coherence blocks imposes the constraint $\tau_\text{p}\leq\tau_\text{c}$, where $\tau_\text{c}$ is the number of symbols (samples) that can be transmitted over one coherence block. Frequently, $K\gg\tau_\text{c}$ makes it impossible to assign mutually orthogonal pilots to all UEs. Therefore, two or more UEs might be assigned the same pilot. We denote the index of the pilot assigned to UE $k$ as $t_k\in\{1,\dots,\tau_\text{p}\}$ and define the set of UEs using the same pilot as UE $k$, including itself:
\begin{equation}
\mathcal{P}_k=\{i: t_i = t_k, i = 1,\dots,K\}\subset\{1,\dots,K\}.
\label{set_of_pilot_sharing_users}
\end{equation}

The elements of $\sqrt{p_k}\boldsymbol{\phi}_{t_k}$ are transmitted by UE $k$ over $\tau_\text{p}$ consecutive samples at the beginning of the coherence block, where $p_k$ is the uplink transmit power of UE $k$. Then, during the entire pilot transmission, the received signal $\textbf{Y}_l^\text{p}\in\mathbb{C}^{N\times\tau_\text{p}}$ at SA $l$ is
\begin{equation}
\textbf{Y}_l^\text{p}=\sum_{i=1}^K\sqrt{p_i}\textbf{h}_{il}\boldsymbol{\phi}_{t_i}^\text{T} + \textbf{N}_l^\text{p},
\label{Y_l}
\end{equation}
where $\textbf{N}_l^\text{p}\in\mathbb{C}^{N\times\tau_\text{p}}$ is the receiver noise with i.i.d. elements distributed as $\mathcal{N}_\mathbb{C}(0,\sigma_\text{n}^2)$.

Besides, the interference coming from non-pilot-sharing UEs can be eliminated, by multiplying the received signal $\textbf{Y}_l^\text{p}$ with the normalized conjugate of the pilot $\boldsymbol{\phi}_{t_k}$, resulting in the processed received pilot signal $\textbf{y}_{{t_k}l}^\text{p}\in\mathbb{C}^N$, which is a sufficient statistic for estimating $\textbf{h}_{kl}$:
\begin{align}
\textbf{y}_{{t_k}l}^\text{p}
= \textbf{Y}_l^\text{p}\boldsymbol{\phi}_{t_k}^* = \sqrt{p_k}\tau_\text{p}\textbf{h}_{kl} + \sum_{i\in\mathcal{P}_k\backslash\{k\}}\sqrt{p_i}\tau_\text{p}\textbf{h}_{il} + \textbf{n}_{{t_k}l}^\text{p},
\label{y_tk_l}
\end{align}
where $\textbf{n}_{{t_k}l}^\text{p}=\textbf{N}_l^\text{p}\boldsymbol{\phi}_{t_k}^*\sim\mathcal{N}_\mathbb{C}(\mathbf{0},\tau_\text{p}\sigma_\text{n}^2\textbf{I}_N)$.

\subsection{Channel estimation}
\label{subsec:ChannelEstimation}
Channel estimates can be obtained based on the observation $\textbf{y}_{{t_k}l}^\text{p}$. The MMSE estimator uses statistical information, {\it i.e.}, the channel means, and the covariance matrices to partially remove the interference from pilot-sharing UEs. Based on the received signal $\textbf{y}_{{t_k}l}^\text{p}$,
the MMSE estimate of $\textbf{h}_{kl}$ is defined as \cite{[2019]Massive_MIMO_With_Spatially_Correlated_Rician_Fading_Channels}:
\begin{equation}
\widehat{\textbf{h}}_{kl} = \overline{\textbf{h}}_{kl} +  \sqrt{p_k}\textbf{R}_{kl}\boldsymbol{\Psi}_{{t_k}l}^{-1}\left(\textbf{y}_{{t_k}l}^\text{p}-\overline{\textbf{y}}_{{t_k}l}^\text{p}\right),
\label{h_NLoS_est_MMSE}
\end{equation}
where
\begin{align}
\boldsymbol{\Psi}_{t_kl} = \frac{1}{\tau_\text{p}}\text{Cov}(\textbf{y}_{t_kl}^\text{p}) = \sum_{i\in\mathcal{P}_k} p_i\tau_\text{p}\textbf{R}_{il}+\sigma_\text{n}^2\textbf{I}_N,
\label{Psi}
\end{align}
and
\begin{equation}
\overline{\textbf{y}}_{{t_k}l}^\text{p} = \sqrt{p_k}\tau_\text{p}\overline{\textbf{h}}_{kl} + \sum_{i\in\mathcal{P}_k\backslash\{k\}}\sqrt{p_i}\tau_\text{p}\overline{\textbf{h}}_{il}.
\end{equation}

The statistical distributions (the mean vector and covariance matrices) are assumed perfectly known, since they are constant during hundreds of channel coherence blocks. In practice, they can be estimated using the sample mean and sample covariance matrices.

The resulting estimation error ${\bm\varepsilon}_{{\bf h}_{kl}}=\textbf{h}_{kl}-\widehat{\textbf{h}}_{kl}$ can be modeled as {a} zero-mean random variable {with} covariance matrix \cite{[2019]Massive_MIMO_With_Spatially_Correlated_Rician_Fading_Channels} given by
\begin{equation}
\textbf{C}_{kl} = \textbf{R}_{kl} - p_k\tau_\text{p}\textbf{R}_{kl}\boldsymbol{\Psi}_{t_kl}^{-1}\textbf{R}_{kl}.
\label{C_kl_MMSE}
\end{equation}

The estimate $\widehat{\textbf{h}}_{kl}$ and the estimation error {$\bm \varepsilon_{{\bf h}_{kl}}$} are independent random variables distributed as\cite{[2019]Massive_MIMO_With_Spatially_Correlated_Rician_Fading_Channels}
\begin{align}
\widehat{\textbf{h}}_{kl}\ & 
\sim \mathcal{N}_\mathbb{C}\left( \overline{\textbf{h}}_{kl}, \textbf{R}_{kl} - \textbf{C}_{kl} \right),
 \\ 
{\bm \varepsilon_{{\bf h}_{kl}}}\
 &  
\sim \mathcal{N}_\mathbb{C}\left( \mathbf{0}_N, \textbf{C}_{kl} \right).
\end{align}

The estimation accuracy is defined by the NMSE of the channel estimate\cite{CFbook}. The NMSE of $\widehat{\textbf{h}}_k$ can be computed as
\begin{align}
\text{NMSE}_k = \frac{\mathbb{E}\{||{\bm \varepsilon_{{\bf h}_{k}}}||^2\}}{\mathbb{E}\{||\textbf{h}_k||^2\}} = \frac{\mathbb{E}\{{\bm \varepsilon_{{\bf h}_{k}}^{\rm H}}\}\mathbb{E}\{{\bm \varepsilon_{{\bf h}_{k}}}\}+\sum\limits_{\substack{l=1}}^L\text{tr}(\textbf{C}_{kl})}{N\sum\limits_{\substack{l=1}}^L\beta_{kl}},
\label{NMSE}
\end{align}
where \colb{${\bm \varepsilon_{{\bf h}_{k}}}=[{\bm \varepsilon_{{\bf h}_{k1}}}^\text{T},\dots,{\bm \varepsilon_{{\bf h}_{kL}}}^\text{T}]^\text{T}$} is the collective channel \colb{estimation error} vector of UE $k$.
\colb{The term $\mathbb{E}\{||{\bm \varepsilon_{{\bf h}_{k}}}||^2\}$ is the mean square error (MSE) of the channel estimation $\widehat{\textbf{h}}_{k}$.}

\subsection{UL Data Transmission}
\label{subsec:DataTransmission}
During the UL data transmission, the $i$th UE transmits the signal $s_i\in\mathbb{C}$, with power $p_i=\mathbb{E}\{|s_i|^2\}$, and the SAs receive a superposition of the signals sent from the $K$ UEs. The received signal $\textbf{y}_l^\text{ul}\in\mathbb{C}^N$ at the $l$th SA is then given by
\begin{equation}
    \textbf{y}_l^\text{ul} = \sum_{i=1}^K \textbf{h}_{il}s_i + \textbf{n}_l^\text{u},
\label{y_l_UL}
\end{equation}
where $\textbf{h}_{il}\in\mathbb{C}^N$ denotes the channel vector between the $k$th UE and the $l$th SA, and $\textbf{n}_l^\text{u}\sim \mathcal{N}_\mathbb{C}(\textbf{0},\sigma_\text{n}^2\textbf{I}_N)$ is the independent additive noise during UL data transmission.

We define the set $\mathcal{M}_k$, containing the SAs that serve the $k$th UE, and the diagonal matrix $\textbf{D}_{kl}\in\mathbb{C}^{N\times N}$, which indicates whether the $l$th SA serves the $k$th UE or not:
\begin{align}
\textbf{D}_{kl} = 
\begin{cases}
\textbf{I}_N,  \quad & \text{if}~ l\in\mathcal{M}_k,\\
\mathbf{0}_N,  \quad& \text{if}~ l\notin\mathcal{M}_k.
\end{cases}
\label{D_kl}
\end{align}

If $\alpha_{kl}=1$ and $\alpha_{kj}=0$, {\it i.e.}, if UE $k$ has LoS to SA $l$ but not to SA $j$, a proper SA selection scheme will very likely give preference to SA $l$ over $j$ in serving UE $k$.

Also, let's define a receive combining vector $\textbf{v}_{kl}\in\mathbb{C}^N$ that SA $l$ could use if it serves UE $k$, and the collective combining vector $\textbf{v}_k=[\textbf{v}_{k1}^\text{T}\dots\textbf{v}_{kL}^\text{T}]^\text{T}$. The collective UL data signal $\textbf{y}^\text{ul}\in\mathbb{C}^{LN}$ is given by
\begin{equation}
\textbf{y}^\text{ul} = 
\begin{bmatrix}
\textbf{y}_1^\text{ul} \\\vdots\\ \textbf{y}_L^\text{ul}
\end{bmatrix}
=
\sum_{i=1}^K\textbf{h}_is_i+\textbf{n},
\label{y_UL}
\end{equation}
where $\textbf{h}_i=[\textbf{h}_{i1}^\text{T}\dots\textbf{h}_{1L}^\text{T}]^\text{T}\in\mathbb{C}^{LN}$ is the $i$th UE collective channel, and $\textbf{n}=[\textbf{n}_1^\text{T}\dots\textbf{n}_l^\text{T}]^\text{T}\in\mathbb{C}^{LN}$ is the collective noise vector along the antennas of all SAs.

In order to obtain an estimate of $s_k$, the CPU computes the inner product between the effective receive combining vector $\textbf{D}_k\textbf{v}_k$, where $\textbf{D}_k=\text{diag}(\textbf{D}_{k1},\dots,\textbf{D}_{kL})$, and the received UL data signals at the serving SAs:
\begin{align}
\widehat{s}_k &= \textbf{v}_k^\text{H}\textbf{D}_k \textbf{y}^\text{ul}\\
&=\textbf{v}_k^\text{H}\textbf{D}_k\widehat{\textbf{h}}_ks_k + \textbf{v}_k^\text{H}\textbf{D}_k{\bm \varepsilon_{{\bf h}_{k}}}s_k + \sum\limits_{\substack{i=1\\i\neq k}}^K \textbf{v}_k^\text{H}\textbf{D}_k\textbf{h}_is_i + \textbf{v}_k^\text{H}\textbf{D}_k\textbf{n}.
\label{s_k_hat_centralized}
\end{align}
Notice that the desired signal $\textbf{v}_k^\text{H}\textbf{D}_k\textbf{h}_ks_k$ was divided into two parts in equation \eqref{s_k_hat_centralized}: the desired signal over the estimated channel, $\textbf{v}_k^\text{H}\textbf{D}_k\widehat{\textbf{h}}_ks_k$, and the desired signal over the unknown channel, $\textbf{v}_k^\text{H}\textbf{D}_k{\bm \varepsilon_{{\bf h}_{k}}} s_k$. Only the first one can be utilized for signal detection. The second one is not useful since only the distribution of the estimation error is known, and thus it is treated as interference or noise when computing the instantaneous effective uplink SINR of UE $k$, given by
\begin{equation}
\text{SINR}_k^\text{ul,c} = 
\frac{p_k|\textbf{v}_k^\text{H}\textbf{D}_k\widehat{\textbf{h}}_k|^2
}{
\sum\limits_{\substack{i=1\\i\neq k}}^K p_i|\textbf{v}_k^\text{H}\textbf{D}_k\widehat{\textbf{h}}_i|^2~+\textbf{v}_k^\text{H}\textbf{Z}_k\textbf{v}_k + \sigma_\text{n}^2||\textbf{D}_k\textbf{v}_k||^2
},
\label{SINR_k_centralized}
\end{equation}
where
\begin{equation}
\textbf{Z}_k = \sum_{i=1}^K p_i\textbf{D}_k\textbf{C}_i\textbf{D}_k.
\label{Z_k}
\end{equation}

Finally, being $\tau_\text{u}/\tau_\text{c}$ the fraction of each coherence block that is used for uplink data transmission, the {\it achievable} SE of UE $k$ in the centralized operation can be expressed as
\begin{equation}
\text{SE}_k^\text{ul,c} = \frac{\tau_\text{u}}{\tau_\text{c}} \mathbb{E}\left\{\text{log}_2\left(1+\text{SINR}_k^\text{ul,c}\right)\right\}.
\label{SE_k_centralized}
\end{equation}

\subsection{Receive Combining}
\label{subsec:ReceiveCombining}

A critical implication of deploying an XL antenna array is that the interference that affects UE $k$ is mainly generated by the UEs that are located at its neighborhood\cite{CFbook}. Thus, we will utilize the {\it partial} MMSE (P-MMSE) for receive combining, defined as:
\begin{equation}
\textbf{v}_k = p_k \left(
\sum_{i\in\mathcal{S}_k}p_i\textbf{D}_k(\widehat{\textbf{h}}_i\widehat{\textbf{h}}_i^\text{H}+\textbf{C}_i)\textbf{D}_k+\sigma_\text{n}^2\textbf{I}_{LN}
\right)^{-1}\textbf{D}_k\widehat{\textbf{h}}_k,
\label{v_k}
\end{equation}
where only the UEs that are served by at least one SA that also serves UE $k$ are considered in the inverse matrix, {\it i.e.}, the UEs that have indices in the set
\begin{equation}
\mathcal{S}_k = \{i:\textbf{D}_k\textbf{D}_i\neq\mathbf{0}_{LN}\}.
\label{S_k}
\end{equation}

\colb{In the special case where $\mathcal{S}_k$ contains the $K$} UEs, the vector $\textbf{v}_k$ becomes the MMSE receive combiner, which is the optimal combiner w.r.t. minimizing the MSE of the estimates $\widehat{s}_k$, and also maximizes the SINR \colb{of the $k$th UE,} as defined in \eqref{SINR_k_centralized}. The P-MMSE combiner does not attain optimal SE, but it is scalable, since both the computation of the combining vector $\textbf{v}_k$ and the estimation of the channel vectors $\textbf{h}_i$, $i\in\mathcal{S}_k$, do not result in a computational complexity that grows without limit when $K\rightarrow\infty$.

\section{Pilot Assignment and Subarray Selection}
\label{sec:PA_SAS}
Depending on the mobility, the size of the normalized coherence block may be smaller than the number of UEs, \textit{i.e.}, $\tau_\text{c}<K$.  The finite length of the coherence block imposes that $\tau_\text{p}<\tau_\text{c}$, such that it is impossible to assign mutually orthogonal pilots to all UEs in overcrowded networks where $K>\tau_\text{c}$, and consequently, $K>\tau_\text{p}$. Furthermore, it may happen that $\tau_\text{p}<K$. Thus, in crowded scenarios, we may need to reuse the $\tau_\text{p}$ mutually orthogonal pilots among the XL-MIMO UEs.

If UEs $k$ and $i$ share the same pilot, {\it i.e.}, $t_k=t_i$, UE $i$ will probably cause much more interference on the estimate $\widehat{s}_k$ than the UEs that use other pilot than $t_k$. If the LSF channel gain $\beta_{kl}$ is stronger than $\beta_{il}$, it may be fortunate to assign to this SA the task of estimating $s_k$ but not $s_i$. Thus, each SA shall estimate the signal of one UE only per pilot sequence, namely the one with the strongest channel, such that each UE will be served by a different subset of SAs. It is a necessary and sufficient condition for scalability since the computational complexity will remain finite as $K\rightarrow\infty$\cite{CFbook}.

However, in some cases it may happen that the $k$th UE channel is not the strongest among the UEs that share the pilot $t_k$ in any SA, such that the CPU would not even process its signal. Thus, the SA selection is insufficient to ensure a uniform quality of service among the XL-MIMO UEs, and a pilot assignment strategy will be required. A possible way to avoid the impairment mentioned above is to assign exclusive orthogonal pilots to the worst channel UEs. Subsection \ref{subsec:PilotAssignment} discusses the proposed pilot assignment strategy.

Both pilot assignment (PA) and subarray selection (SAS) strategies are based only on the channel statistics, not on the instantaneous channel realizations; hence, we do not need to run both algorithms in every coherence block. A basic algorithm for SA selection is to first assign pilots to all UEs and then let each SA serve only one UE per pilot, namely the one with the strongest LSF channel gain\footnote{Recall that this is a channel statistical information, which has been assumed known. Indeed, t is assumed that the algorithms can count on the perfect knowledge of the channel statistics (channel covariance matrix and the channel mean) since they change much slower than the channel realizations.} among the subset of UEs that have been assigned that pilot. Note that each SA can make these decisions locally \cite{CFbook}.

\subsection{Pilot Contamination Effect}
\label{subsec:PilotContamination}
Suppose that UEs $k$ and $i$ use the same pilot. Intuitively, the closer the UE $i$ is to the antenna array, the higher its NLoS LSF channel gain ($\beta_{il}^\text{NLoS}$) will be and, therefore, the more it will interfere with the estimation of $\textbf{h}_{kl}$. It can be verified in \eqref{C_kl_MMSE} and \eqref{NMSE} that the NMSE increases with $\beta_{il}^\text{NLoS}$, which is implicitly included in $\boldsymbol{\Psi}_{{t_k}l}^{-1}$. Furthermore, if UE $k$ is close to the ULA, and especially if there is a LoS path, the associated channel will be stronger, and the relative estimation error will be smaller since the LoS component changes slowly; therefore, it is assumed to be perfectly known. On the other hand, yet according to \eqref{NMSE}, the NMSE of estimating $\textbf{h}_k$ is inversely proportional to $\sum_{l=1}^L\beta_{kl}$.

Moreover, if $\textbf{R}_{kl}$ is invertible, one can write the relation\cite{CFbook}
\begin{equation}
\widehat{\textbf{h}}_{il} = \overline{\textbf{h}}_{il} + \sqrt{\frac{p_i}{p_k}}\textbf{R}_{il}\textbf{R}_{kl}^{-1}(\widehat{\textbf{h}}_{kl} - \overline{\textbf{h}}_{kl}).
\label{hil_hkl}
\end{equation}

Notice that the more similar the matrices $\textbf{R}_{kl}$ and $\textbf{R}_{il}$ are, the more correlated the estimates $\widehat{\textbf{h}}_{kl}$ and $\widehat{\textbf{h}}_{il}$ will be. Although the channels $\textbf{h}_{kl}$ and $\textbf{h}_{il}$ are statistically independent, the MMSE estimates will be fully correlated if the NLoS channels are spatially uncorrelated, {\it i.e.}, if $\textbf{R}_{kl}=\beta_{kl}^\text{NLoS}\textbf{I}_N$ and $\textbf{R}_{il}=\beta_{il}^\text{NLoS}\textbf{I}_N$, since they will be the same except for a scaling factor\cite{CFbook}.

Since two closely located UEs, $k$ and $i$, have approximately the same azimuth and elevation angles, they have similar spatial correlation properties, {\it i.e.}, the matrices $\textbf{R}_{kl}$ and $\textbf{R}_{il}$ can result quite similar. As a result, if they use the same pilot, the MMSE estimates of $\textbf{h}_{kl}$ and $\textbf{h}_{il}$ will be correlated, which was demonstrated analytically and numerically in \cite{CFbook} to cause strong pilot contamination.


\subsection{Pilot Assignment (PA) in XL-MIMO}
\label{subsec:PilotAssignment}

The pilot assignment (PA) is a combinatorial problem with $\tau_\text{p}^K$ possible assignments and it is unfeasible to evaluate all of them in realistic 5G mMIMO network scenarios\cite{CFbook}. Hence, a practical, suboptimal PA algorithm typically assigns pilots to the UEs iteratively. The following PA techniques have been considered in this work.

\vspace{2mm}
\noindent\textbf{\textit{Random PA}}: is the most naive strategy, consisting of assigning pilots randomly to all the $K$ UEs. It does not guarantee \colb{any pilot contamination control} but has zero computational complexity.

\vspace{2mm}
\noindent\textbf{\textit{Genie PA}}: an exhaustive search can be used to obtain the best PA solution concerning an optimization criterion, for instance, the average channel estimation NMSE among the $K$ UEs. The closed-form expression given in \eqref{NMSE} can be used as a cost function for evaluating each candidate solution. \colb{The NMSE can be evaluated through the closed-form expression in \eqref{NMSE}, such that the cost function is given by}
\begin{equation}
\colb{
\sum_{k=1}^K\frac{\sum\limits_{\substack{l=1}}^L\text{tr}(\textbf{C}_{kl})}{N\sum\limits_{\substack{l=1}}^L\beta_{kl}}
}
\label{cost_function}
\end{equation}
\colb{where the term ${\bm \varepsilon_{{\bf h}_{kl}}}$ in \eqref{NMSE} was removed since we consider MMSE channel estimator, which is unbiased.} The genie PA procedure is shown in Algorithm \ref{alg:PA:genie}. However, this method is feasible only if the number of possible assignments is small. \colb{It guarantees optimality, but it is unfeasible since there are $(\tau_\text{p})^K$ possible assignments\cite{CFbook}.}

\begin{algorithm}[hbt]
\caption{\bf Genie Pilot Assignment}
\label{alg:PA:genie}
\textbf{{Input}: $\textbf{R}_{kl}$, $\forall kl$.}
\begin{algorithmic}[1]
\STATE {Initialize the population $\boldsymbol{\Theta}_\text{G}$ containing in its columns all the $A_\text{G}$ possible PA candidates.}
\FOR{{$i=1,\dots,\textbf{K}$}}
    \STATE {Using equation \eqref{NMSE}, compute $\frac{1}{N}\sum\limits_{\substack{k=1}}^K\text{NMSE}_k$ as if the pilots were assigned according to the $i$th column of the matrix $\boldsymbol{\Theta}_\text{G}$, and assign it to the $i$th column of the cost vector $\textbf{c}_\text{G}$.}
\ENDFOR
\STATE {Find the pilots $t_1,\dots,t_K$ corresponding to the column of $\boldsymbol{\Theta}_\text{G}$ associated to the smallest element of $\textbf{c}_\text{G}$.}
\end{algorithmic}
\textbf{Output:} pilot assignment $t_1,\dots,t_K$
\end{algorithm}

\vspace{2mm}
\noindent\textbf{\textit{Greedy PA}}: is a simple PA strategy consisting in assigning the $\tau_\text{p}$ orthogonal pilots to $\tau_\text{p}$ random UEs, in the first step, and then assigning pilots to the remaining UEs, one after the other, as to minimize pilot contamination, in the second step. The pilot contamination faced by a UE $k$ when assigning to it a pilot $t$ can be quantified as the sum of the NLoS LSF channel gains of the UEs that were already assigned pilot $t$. It can be considered only the LSF channel gains to SA $l_k$, which is defined as the SA to which the UE $k$ has the strongest LSF channel gain, as in Algorithm \ref{alg:PA:greedy}, or the sum LSF channel gains among the $L$ SAs. \colb{The greedy algorithm comes with no performance guarantee, but effectively avoids closely-located UEs from being assigned the same pilots\cite{CFbook}.}

\begin{algorithm}[hbt]
\caption{{\bf Greedy Pilot Assignment}}
\label{alg:PA:greedy}
\textbf{Input}: LSF coefficients
\begin{algorithmic}[1]
\FOR{$k=1,\dots,\tau_\text{p}$}
    \STATE $t_k \leftarrow k$
\ENDFOR
\FOR{$k=\tau_\text{p},\dots,K$}
    \STATE 
    $l_k \leftarrow \underset{l\in \{1,\dots,L\}}{\text{arg~max}}\beta_{kl}$
    \STATE $\tau \leftarrow \underset{t\in \{1,\dots,\tau_\text{p}\}}{\text{arg~min}}\sum\limits_{\substack{i=1\\t_i=t}}^{k-1}\beta_{il_k}^\text{NLoS}$
    \STATE $t_k\leftarrow \tau$
\ENDFOR
\end{algorithmic}
\textbf{Output:} pilot assignment $t_1,\dots,t_K$
\end{algorithm}

\vspace{2mm}
\noindent\textbf{\textit{Genetic Algorithm (GA) PA}}. Herein we propose to solve the PA problem in XL-MIMO networks by applying heuristic genetic algorithm (GA). It uses $A$ random PA candidates per iteration. \colb{The candidates are $K$-length vectors, where the $k$th entry is the index of the pilot assigned to the $k$th UE.} Each candidate is evaluated in terms of a cost function\colb{. Here, for a fair comparison, we will take the same cost function used by the genie algorithm (exhaustive search), given by \eqref{cost_function}, {\it i.e.}, the average channel estimation NMSE among the $K$ UEs.}

The $\phi$ best candidates in the population are used to generate the $A$ members of a new population. \colb{From the list of the best candidates, $A$ pairs of candidates are picked at random to generate the $A$ descendants (elements of the population in the next iteration of the algorithm). The descendants are generated by applying the crossover operator between their correspondent parents, with a random crossover point that is comprised in the interval $[2,\tau_\text{p}]$.} Then, the mutation operator is applied to modify the pilot of each UE. The probability of modification is $p_\text{mut}$, \colb{which means that the probability of UE $k$ pilot changing from $t_k$ to another pilot is $p_\text{mut}$. Specifically, the probability of changing from $t_k$ to each of the other $\tau_\text{p}$ is the same.} At the end of each iteration, the best candidate is updated if it is better than the previous solution (at the previous iteration). The predefined number of iterations is denoted by $N_\text{it}$. The pseudo-code for the heuristic GA-based PA method is presented in Algorithm \ref{alg:PA:GA}.

\begin{algorithm}[!hbtp]
\caption{\bf Pilot Assignment via Genetic Algorithm}
\label{alg:PA:GA}
\textbf{Input}: $A$, $\phi$, $p_\text{mut}$, $N_\text{it}$, $\textbf{R}_{kl}$, $\forall kl$.
\begin{algorithmic}[1]
\STATE {Initialize the population $\boldsymbol{\Theta}$ with $A$ random candidates.}
\STATE {Evaluate the cost of each candidate in $\boldsymbol{\Theta}$, according to \colb{\eqref{NMSE}}, forming the vector $\textbf{c}$.}
\STATE {Sort $\textbf{c}$ in ascending order, reorganizing the columns of $\boldsymbol{\Theta}$ accordingly.}
\STATE {Initialize $c_\text{best}=c_1$, and $\boldsymbol{\theta}_\text{best}=\boldsymbol{\theta}_1$.}
\FOR{{$i=2,\dots,N_\text{it}$}}
    \FOR{{$a=1,\dots,A$}}
        \STATE Select randomly two of the $\phi$ best candidates in the population $\boldsymbol{\Theta}$ to be the parents of $\boldsymbol{\theta}_a$;
         \STATE Apply the crossover operator in a random crossover point $\in\{2,K\}$.
        \STATE {Apply the mutation operator in $\boldsymbol{\theta}_a$ with probability $p_\text{mut}$.}
        \STATE {Evaluate the cost function of $\boldsymbol{\theta}_a$\colb{, according to \colb{\eqref{NMSE}},} and assign it to $c_a$.}
    \ENDFOR
    \STATE {Sort $\textbf{c}$ in descending order, reorganizing the columns of $\boldsymbol{\Theta}$ accordingly.}
    \IF{{$c_1>c_\text{best}$}}
        \STATE {Update $c_\text{best}=c_1$ and $\boldsymbol{\theta}_\text{GA}=\boldsymbol{\theta}_1$.}
    \ENDIF
\ENDFOR
\end{algorithmic}
{\textbf{Output:} $\boldsymbol{\theta}_\text{GA}$}
\end{algorithm}

After assigning the pilots, the CPU must inform each SA on the designated pilot for each UE covered by the specific SA. This is not significant in terms of fronthaul traffic since the pilot decisions change according to the channel statistics, not on the instantaneous channels basis.

\colb{The proposed GA-based PA for XL-MIMO systems is a suboptimal algorithm since it is an iterative method searching for solutions in a very reduced  search space, capable to approach the {\it genie} solution, which uses the entire search space. Moreover, the heuristic evolutionary GA-based pilot assignment method has the potential to be much more successful than {\it greedy} PA -- and mainly random-based searching methods -- in reducing pilot contamination since the former aggregates some intelligence in selecting potential candidates/solution beyond the simple cost function evaluation.}



\subsection{Subarray \colb{selection} in XL-MIMO}
\label{sec:SubarraySelection}
The SA selection is tightly connected to the pilot assignment procedure, and will be done after all the UEs have been assigned to pilots. For that, in Algorithm \ref{alg:SAS}, as in \cite[Algorithm 4.1]{CFbook}, each SA serves one and only one UE per pilot, namely the one with the largest LSF coefficient among the UEs using that pilot. It makes good practical sense, since each SA can decide locally which UEs to serve. If the pilot assignment is performed properly, it is highly probable that every UE will be served by at least one SA.



\begin{algorithm}[hbt]
\caption{{\bf {Strongest-UE SA \colb{Selection}}}}
\label{alg:SAS}
\textbf{Input}: LSF coefficients and pilot assignment $t_1,\dots,t_K$
\begin{algorithmic}[1]
\STATE \textbf{Initialization:} Set $\mathcal{M}_1=\dots=\mathcal{M}_K=\emptyset$.
\FOR{$l=1,\dots,L$}
\FOR{$t=1,\dots,\tau_\text{p}$}
\STATE $k\leftarrow \underset{i\in\{1,\dots,K\};t_i=t}{\text{arg~max}} \beta_{il}$
\STATE $\mathcal{M}_k\leftarrow\mathcal{M}_k\cup\{l\}$
\ENDFOR
\ENDFOR
\end{algorithmic}
\textbf{Output:} $\mathcal{M}_1=\dots=\mathcal{M}_K$
\end{algorithm}
	
\section{Numerical Results}
\label{sec:Results}


In this section, we present numerical results based on Monte Carlo simulations (MCS) to show the influence of the different proposed pilot assignment strategies and channel estimators on the estimation accuracy and the SE in XL-MIMO systems. The antenna array containing $L$ uniformly spaced SAs is centered in the origin of the Cartesian plane and has a length of 100 m, extending along the y-axis, as depicted in Fig. \ref{fig:angles_and_distances}. Each SA contains $N=4$ antennas, spaced by a distance of half a wavelength. In each MCS realization, the $K$ UEs are randomly dropped on a rectangular area, extending in the range $[\pm 100]$ m, both on the x-axis and on the y-axis. Table \ref{tab:parameters} contains the adopted values for the simulation parameters.

\begin{table}[!htbp]
\caption{Adopted Simulation Parameters values}
\begin{center}
\begin{tabular}{|l|c|} \hline
\textbf{Parameter} & \textbf{Value} \\ \hline \hline
\colb{Number of subarrays: $L$} & \colb{\{25,50\}} \\ \hline
\colb{Number of antennas per subarray: $N$} & \colb{4} \\ \hline
Height of the antenna array: $z_l^\text{SA}$ & 10 m \\ \hline
	Height of the UEs: $z_l^\text{UE}$ & 1.5 m\\ \hline
	NLoS path-loss attenuation exponent: $\gamma$ & 4\\ \hline
	Shadowing standard deviation: $\sigma_\text{s}^2$ & 3 dB (LoS)\\ & 4 dB (NLoS)\\ \hline
	Azimuth angle standard deviation: $\sigma_\varphi$ & $10^\circ$\\ \hline
	Elevation angle standard deviation: $\sigma_\theta$ & $10^\circ$\\ \hline
	Median channel gain at a distance of 1 m: $\beta_0$ & $8.9125\cdot10^{-4}$\\ \hline
	Array length & 100 m\\ \hline
    Wavelength: $\lambda$ & 12.5 cm \\ \hline
    \colb{Separation distance between consecutive antennas in the same SA}: $d$ & $\lambda/2$\\ \hline
    Number of antennas in each SA: $N$ & 4\\ \hline
    LoS Probability between UE$_k$--SA$_l${: $\text{Pr}(\alpha_{kl}=1)$} & Eq. \eqref{q}\\  
    \hline
    UE transmit power: $p_1,\dots,p_K$ & 10 dBm\\ \hline
    Noise power: $\sigma_\text{n}^2$ & $-96$ dBm\\ \hline
    Pilot sequences' length: $\tau_\text{p}$ & 4\\ \hline
    Number of Monte-Carlo realizations & 1000 \\ \hline
\end{tabular}
\end{center}
\label{tab:parameters}
\end{table}

The GA-based PA algorithm was also configured to approach the optimal solution in minimizing the average NMSE.  The stopping criterion is based on the number of iterations $N_{\rm it}$ and the number of candidates in the population is set as twice the number of UEs; the probability of mutation is adjusted to $p_{\rm mut}=0.02$ and the number of best candidates is set to be constrained by $A/2$. 
The adopted parameter values in the GA-based PA algorithm are listed in Table \ref{tab:parameters_GA}. \colb{The proposed GA-based pilot allocation method is compared with three different procedures: a) genie-PA, b) greedy-PA, and c) random-PA methods. SA selection step
is applied equally for the four PA methods.}

\begin{table}[H]
\caption{Adopted parameters values for simulating the GA-based PA algorithm}
\begin{center}
\small
\begin{tabular}{|l|c|} \hline
	\textbf{Parameter} & \textbf{Value} \\ \hline \hline
	Number of iterations: $N_\text{it}$ & 15 \\ \hline
	Number of candidates in the population: $A$ & $2K$ \\ \hline
	Probability of mutation: $p_\text{mut}$ & 2\% \\ \hline
	Number of best candidates: $\phi$ & $\lceil A/2 \rceil$ \\ \hline
\end{tabular}
\end{center}
\label{tab:parameters_GA}
\end{table}

For the micro-urban scenario, the probability of existing LoS between UE $k$ and SA $l$ is a function of the distance between them and is modeled according to \cite{3GPP.TR.36.814}  as:
\begin{align}
\text{Pr}(\alpha_{kl}=1) = \text{min}\left(\frac{18~\text{m}}{d_{kl}},\,1\right)\cdot\left[1-e^{-\frac{d_{kl}}{36~\text{m}}}\right]+e^{-\frac{d_{kl}}{36~\text{m}}}.
\label{q}
\end{align}

According to \eqref{q}, the propagation is always in LoS conditions for $d_{kl}\leq18~\text{m}$, {\it i.e.}, $q = 1$, while, for $d_{kl}>18~\text{m}$, the LoS probability decays exponentially. Considering the LoS probability given by equation \eqref{q} and the adopted simulation parameters for array length and cell dimensions, we obtained an average LoS probability of 36\%. It represents an expected condition of the XL-MIMO system operating in micro-urban channel scenarios. From the trustworthy numerical simulations perspective, we cannot disregard the LoS component in urban crowded channel scenarios since most UEs are close to the BS array, which is much longer than co-located mMIMO arrays.

\subsection{Achievable Spectral Efficiency}

\colb{We compare the channel estimation NMSE and the per-user SE obtained with the proposed GA-based PA strategy and the other three strategies mentioned in Section \ref{sec:PA_SAS}: random, greedy, and genie PA methods. According to Fig. \ref{fig:fluxogram}, the system operation is composed by the following steps: pilot assignment, SA selection (Alg. \ref{alg:SAS}), channel estimation and receive combining.}

\begin{figure}[htbp!]
\centering
\includegraphics[trim={0mm 0mm 0mm 0mm} ,clip,width=.5\linewidth]{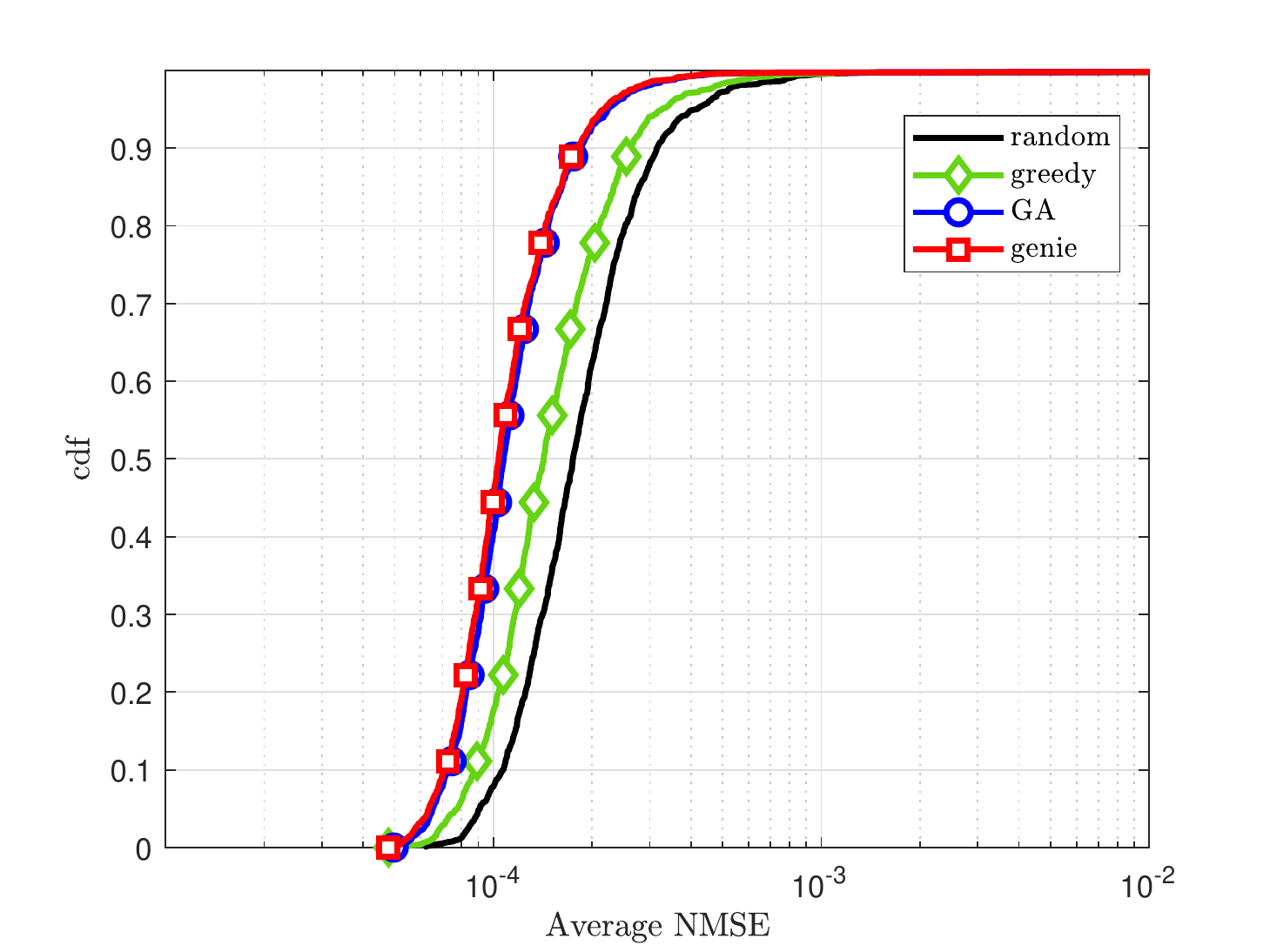}
\vspace{-4mm}
\caption{\colb{Cdf of the average channel estimation NMSE, among the $K$ UEs.}}
\label{fig:R1:avgNMSE}
\end{figure}

\colb{First, recall that the task of both the genie and the GA-based PA algorithms is to find the pilot assignment solution that minimizes the {\it average} channel estimation NMSE among the $K$ UEs; the first does it via exhaustive search, while the second is a GA approach. Fig. \ref{fig:R1:avgNMSE} shows the cumulative density function (cdf) of the average channel estimation NMSE, {\it i.e.}, $\overline{\text{NMSE}}=\frac{1}{K}\sum_{k=1}^K\text{NMSE}_k$, where $\text{NMSE}_k$ is computed via eq. \eqref{NMSE}. The simulation setup assumes $K=6$, $\tau_\text{p}=3$, $L=25$ and $N=4$. According to Fig. \ref{fig:R1:avgNMSE}, simulations confirmed that genie and GA strategies actually provide the smallest average NMSE values in every channel realization; one can also verify that the GA-based PA strategy is efficient since it achieves approximately the same performance as the genie PA strategy but requires less computation (reduced search space).}

\begin{figure}[htbp!]
\centering
\subfloat[\colb{Maximum channel estimation NMSE among the $K$ UEs}]{\label{fig:R1:maxNMSE}{\includegraphics[width=0.495\textwidth]{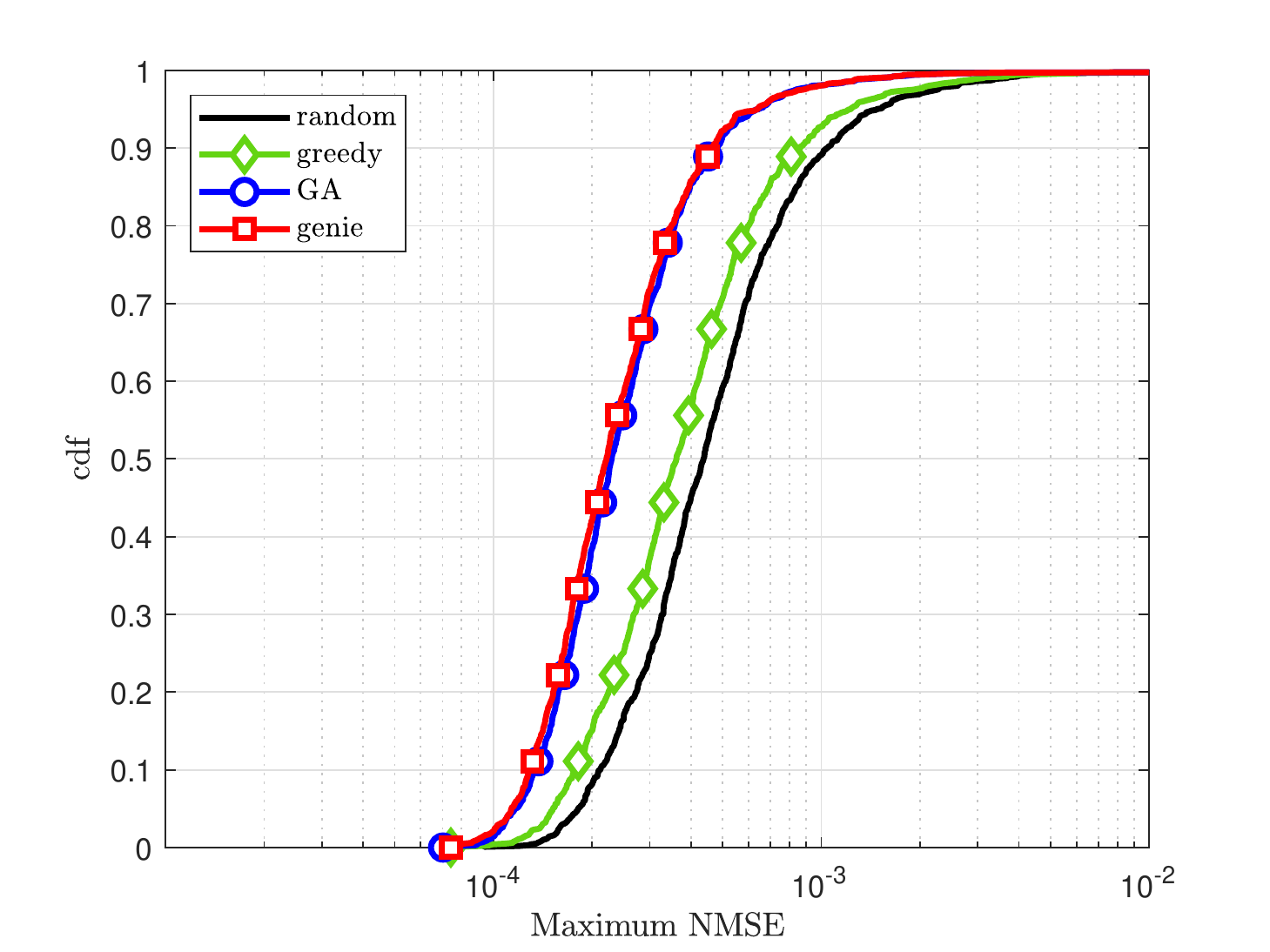}}}\hfill
\subfloat[\colb{Minimum per-user SE}]{\label{fig:R1:min_per_user_SE}{\includegraphics[width=0.495\textwidth]{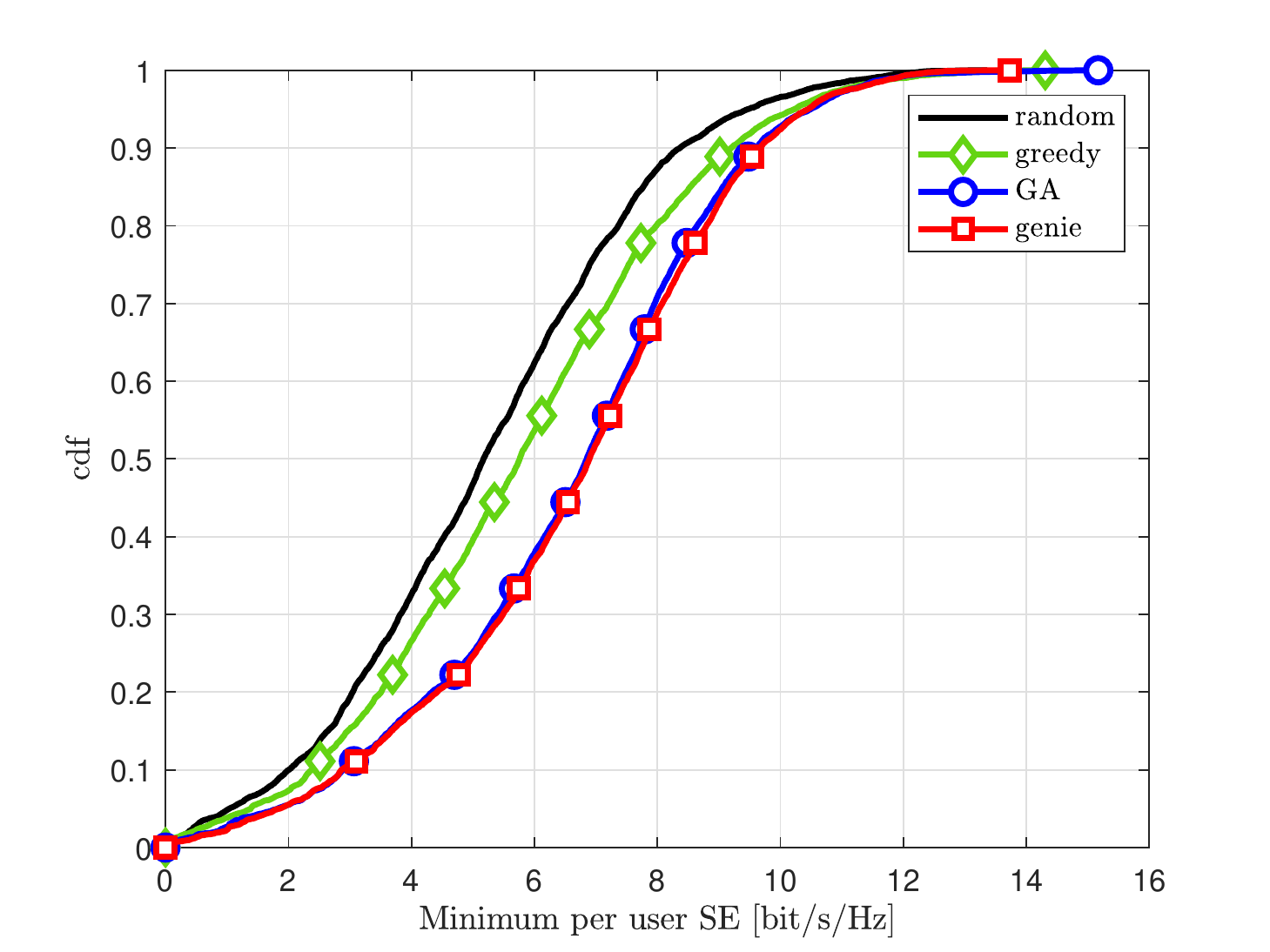}}}\hfill
\subfloat[\colb{Minimum channel estimation NMSE among the $K$ UEs}]{\label{fig:R1:minNMSE}{\includegraphics[width=0.495\textwidth]{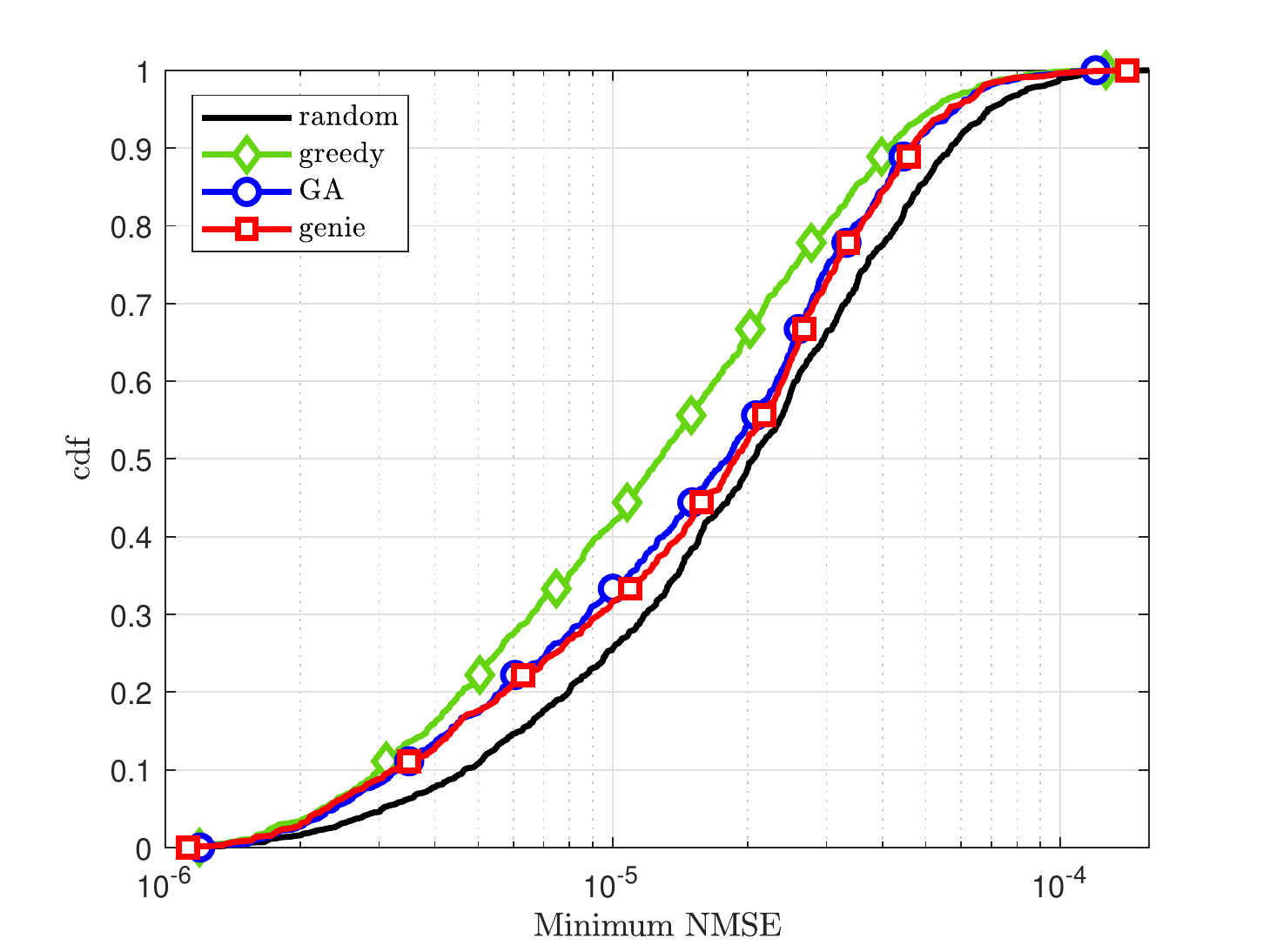}}}\hfill
\subfloat[\colb{Maximum per-user SE}]{\label{fig:R1:max_per_user_SE}{\includegraphics[width=0.495\textwidth]{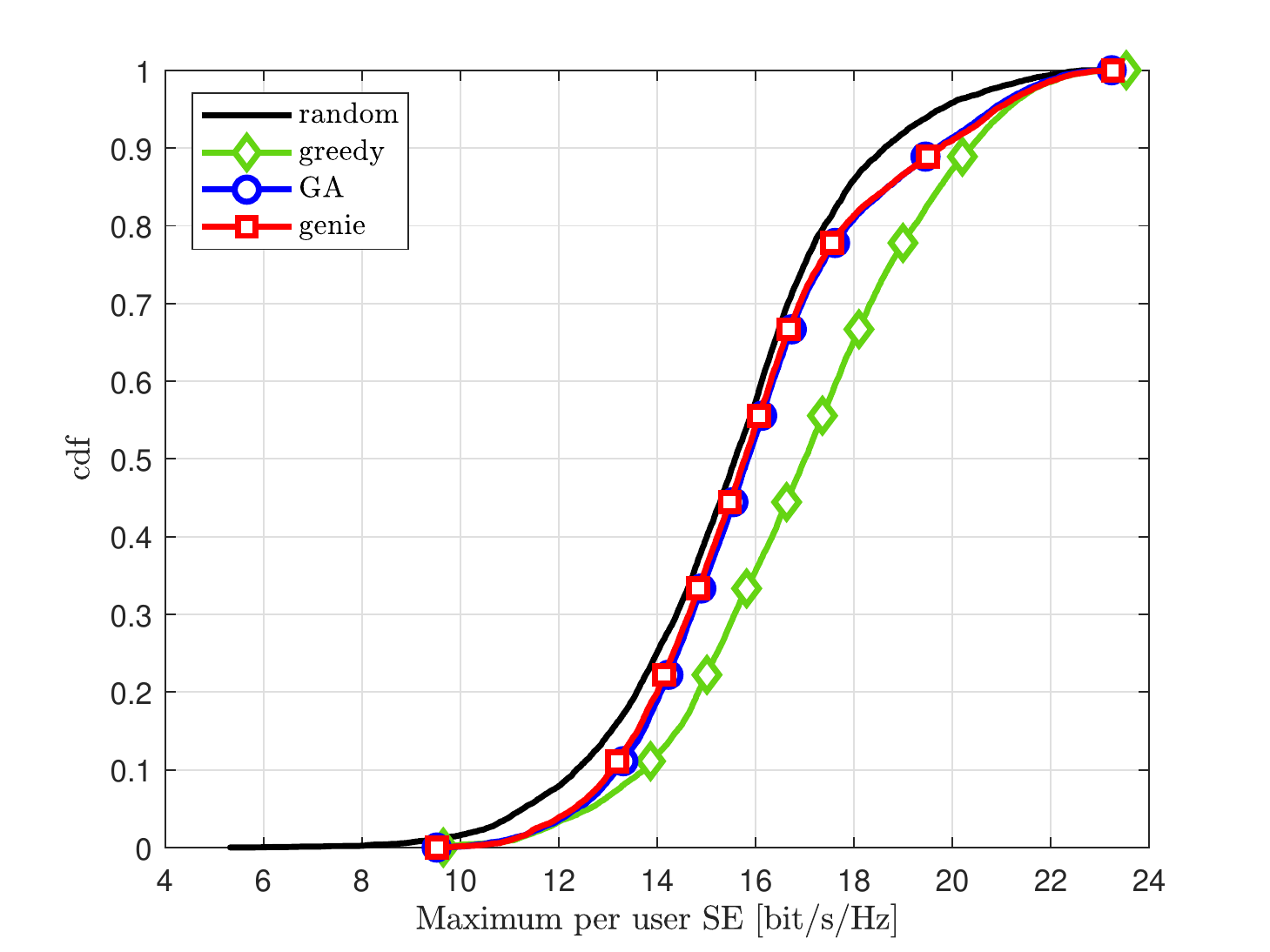}}}\hfill
\subfloat[\colb{Channel estimation NMSE ($\text{NMSE}_k$)}]{\label{fig:R1:NMSE}{\includegraphics[width=0.495\textwidth]{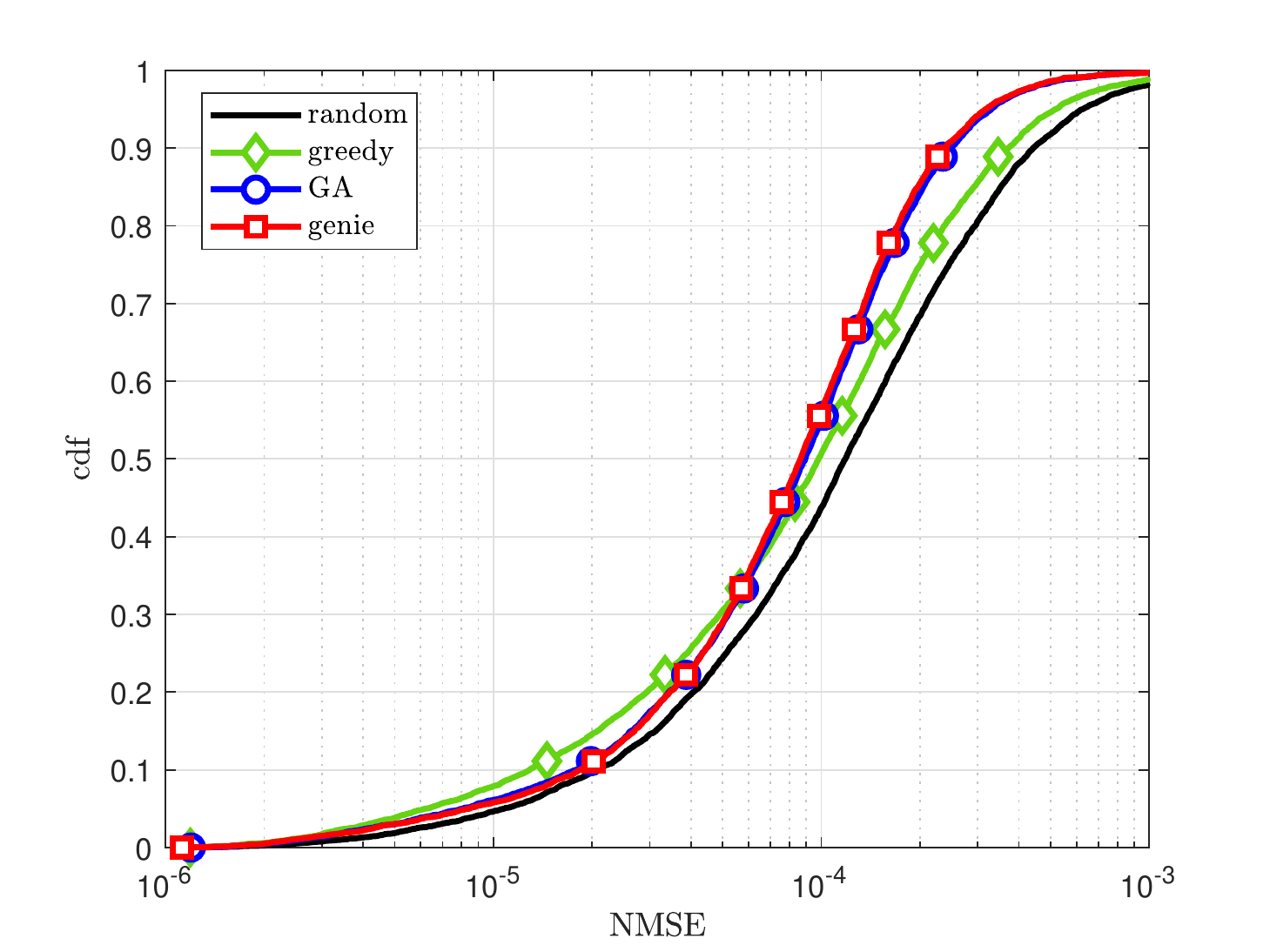}}}\hfill
\subfloat[\colb{Per-user SE}]{\label{fig:R1:per_user_SE}{\includegraphics[width=0.495\textwidth]{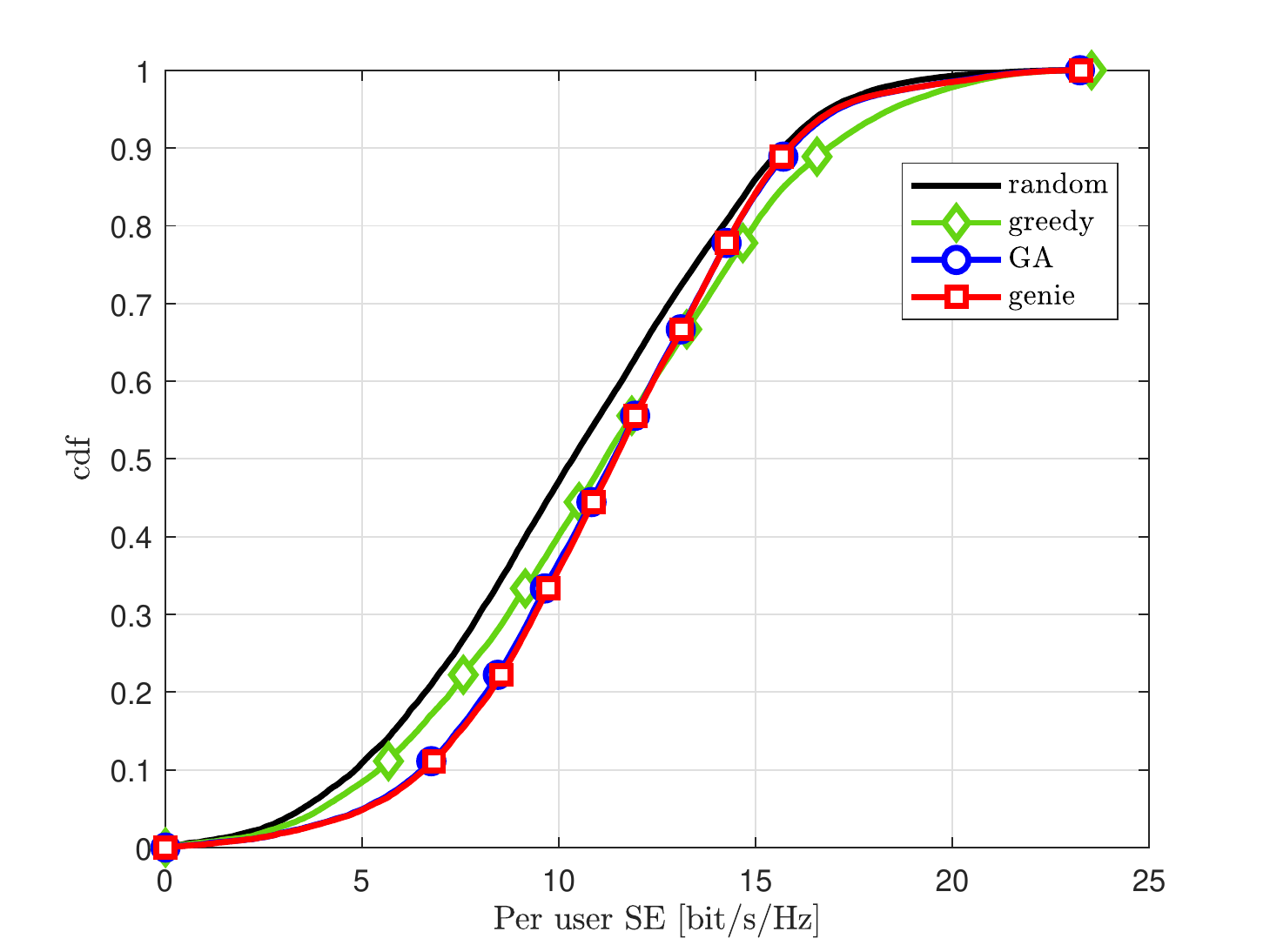}}}\hfill
\caption{\colb{Cdf of the channel estimation NMSE and the per-user SE, considering the worst UE in each Monte-Carlo channel realization (a and b), the best UE (c and d) and all the $K$ UEs (e and f), with $K=6$, $\tau_\text{p}=3$, $L=25$ and $N=4$.}} 
\label{fig:R1:NMSE_SE}
\end{figure}

\colb{Figs. \ref{fig:R1:maxNMSE} and \ref{fig:R1:min_per_user_SE} present, respectively, the cdf of the maximum per user channel estimation NMSE and the cdf of the minimum per user SE, among the $K$ UEs that are simultaneously served by the XL-MIMO array, representing the worst case in each channel realization, among the $K$ UEs. According to Fig. \ref{fig:R1:min_per_user_SE}, the genie PA strategy, which is simply an exhaustive search for the PA solution that minimizes the average channel estimation NMSE among the $K$ UEs, achieved the best result regarding the maximization of the minimum per user SE among the $K$ UEs. The GA-based PA algorithm achieved almost the same performance, since this algorithm repeatedly modifies a population, selecting the best ones in terms of the same optimization criterion as genie PA, {\it i.e.}, the minimization of the average channel estimation NMSE, but using a reduced search space.}

\begin{figure}[htbp!]
\centering
\includegraphics[trim={0mm 0mm 0mm 0mm} ,clip,width=.5\linewidth]{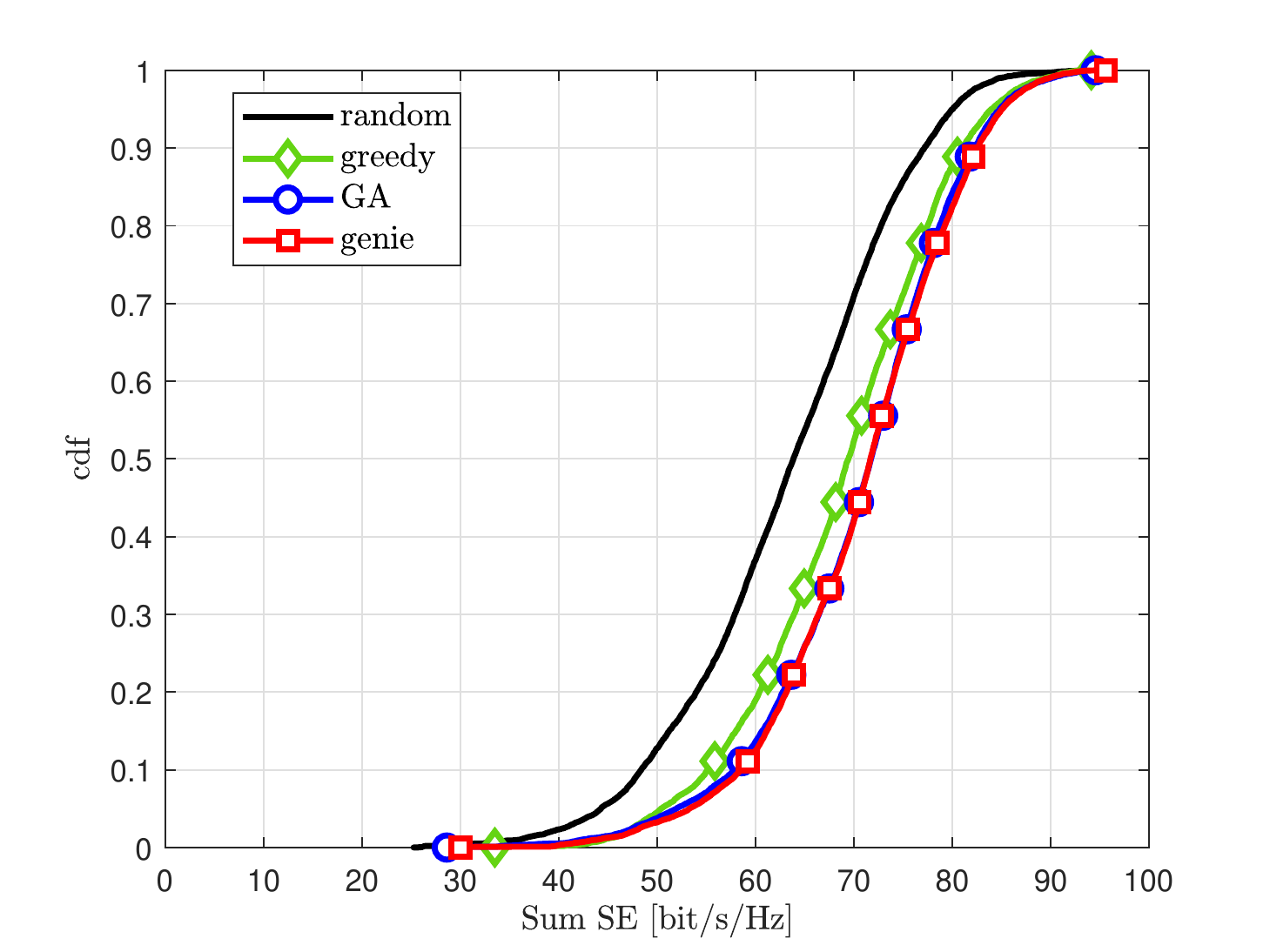}
\vspace{-4mm}
\caption{\colb{Cdf of the sum SE.}}
\label{fig:R1:sumSE}
\end{figure}

\colb{Furthermore, one can see that the proposed GA-based PA beats both random and greedy PA in all channel realizations, which means that it is significantly improving the performance of the UEs with bad channel conditions, compared with these two methods. On the other hand, the proposed method is inferior to the greedy strategy for the UEs with very good channel conditions, according to Figures \ref{fig:R1:minNMSE} and \ref{fig:R1:max_per_user_SE}, which presents respectively the cdf of the minimum per user channel estimation NMSE and the cdf of the maximum per user SE, among the $K$ UEs. Finally, Figures \ref{fig:R1:NMSE} and \ref{fig:R1:per_user_SE} presents, respectively, the cdf of the per user channel estimation NMSE and the cdf of the per user SE, considering all the $K$ UEs in each Monte-Carlo channel realization. Again, one can observe that the GA-based PA outperforms the random and greedy strategies, except for the UEs with good channel conditions. The achievements of the proposed PA strategy for XL-MIMO are in line with the requirements of beyond 5G networks, where making the quality of service more uniform is more important than improving the peak data rates. Finally, Figure \ref{fig:R1:sumSE} shows that the proposed PA strategy also provides higher sum rates than the random and the greedy strategies. The simulation setup of Figs. \ref{fig:R1:NMSE_SE} and \ref{fig:R1:sumSE} is $K=6$, $\tau_\text{p}=3$, $L=25$ and $N=4$.}

\colb{Elaborating further on the achieved SE, one can infer that more accurate channel estimates leads to higher SEs. When computing the average of the channel estimation NMSE among a set of UEs, the worst channel UEs have much more impact on the average NMSE than the UEs with good or moderate channel condition. Thus, when pilots are suitably assigned aiming to minimize the average channel NMSE estimation, which is the case of GA-based and genie PA schemes, we are moving towards increasing mainly the SE of the worst channel UEs, which is observed in \ref{fig:R1:NMSE_SE}.}

\begin{figure}[htbp!]
\centering
\subfloat[Average sum SE]{\label{fig:sumSExK}{\includegraphics[width=0.495\textwidth]{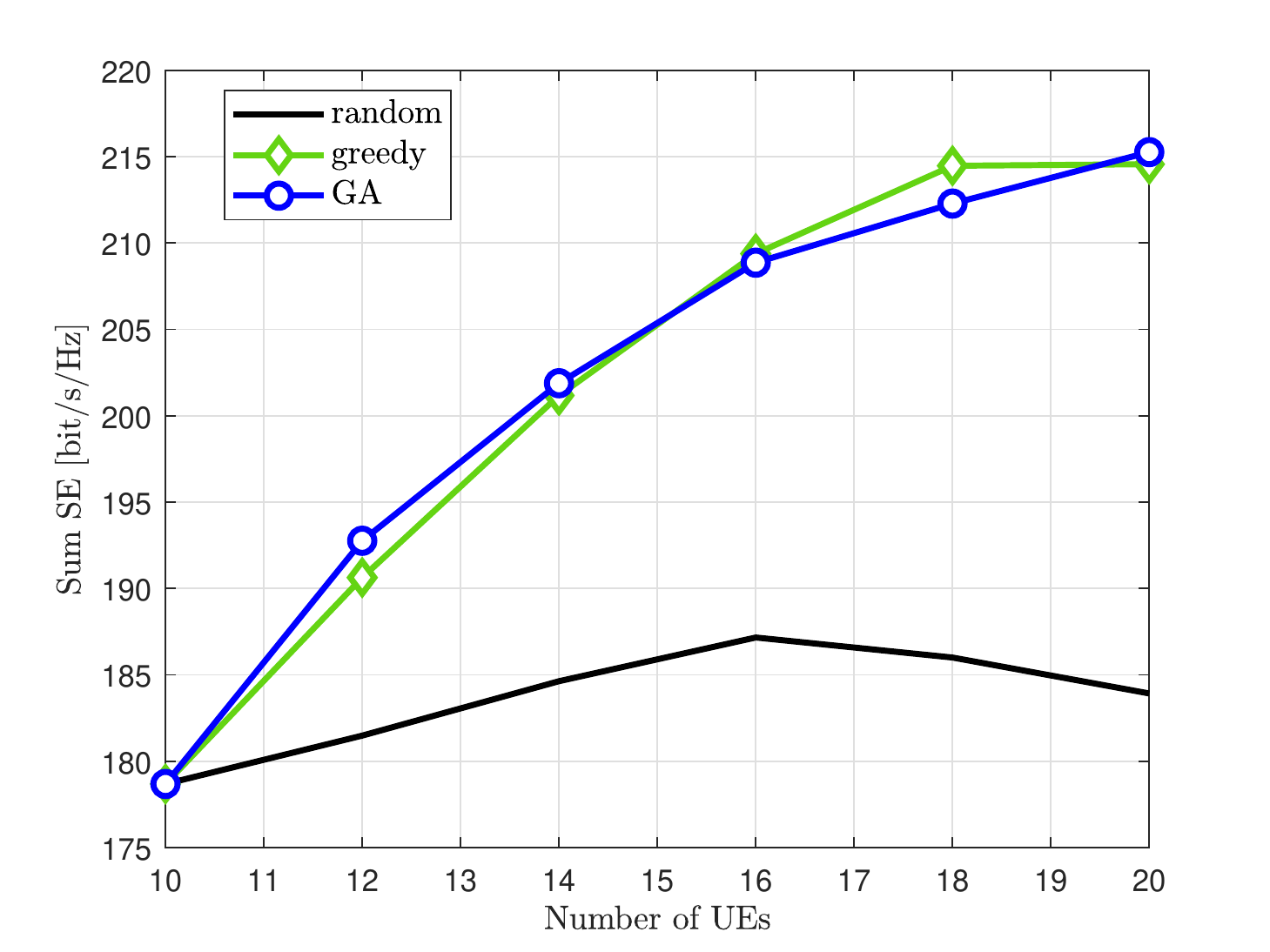}}}\hfill
\subfloat[Average per user SE]{\label{fig:peruserSExK}{\includegraphics[width=0.495\textwidth]{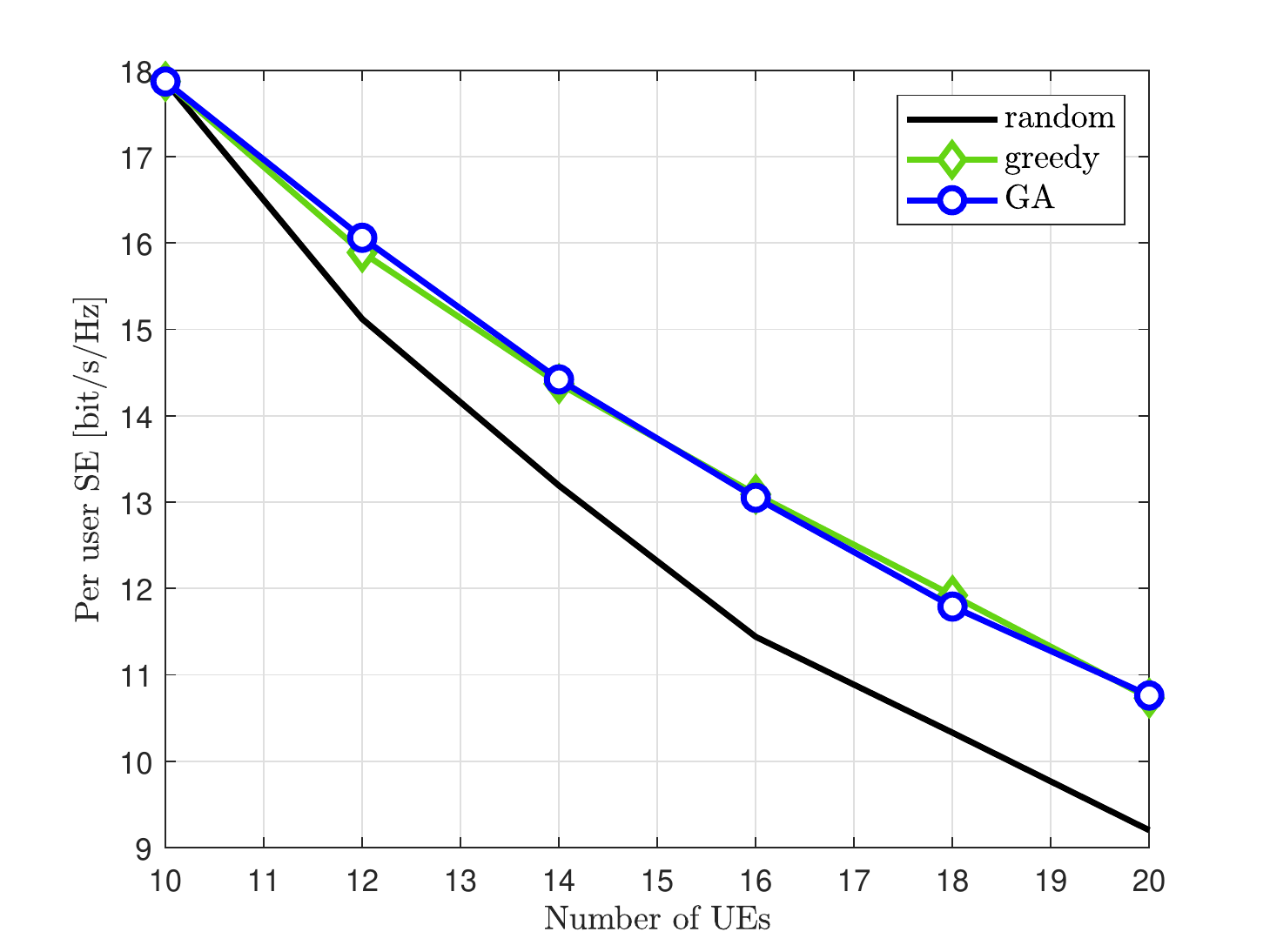}}}\hfill
\subfloat[Average minimum per user SE]{\label{fig:minSExK}{\includegraphics[width=0.495\textwidth]{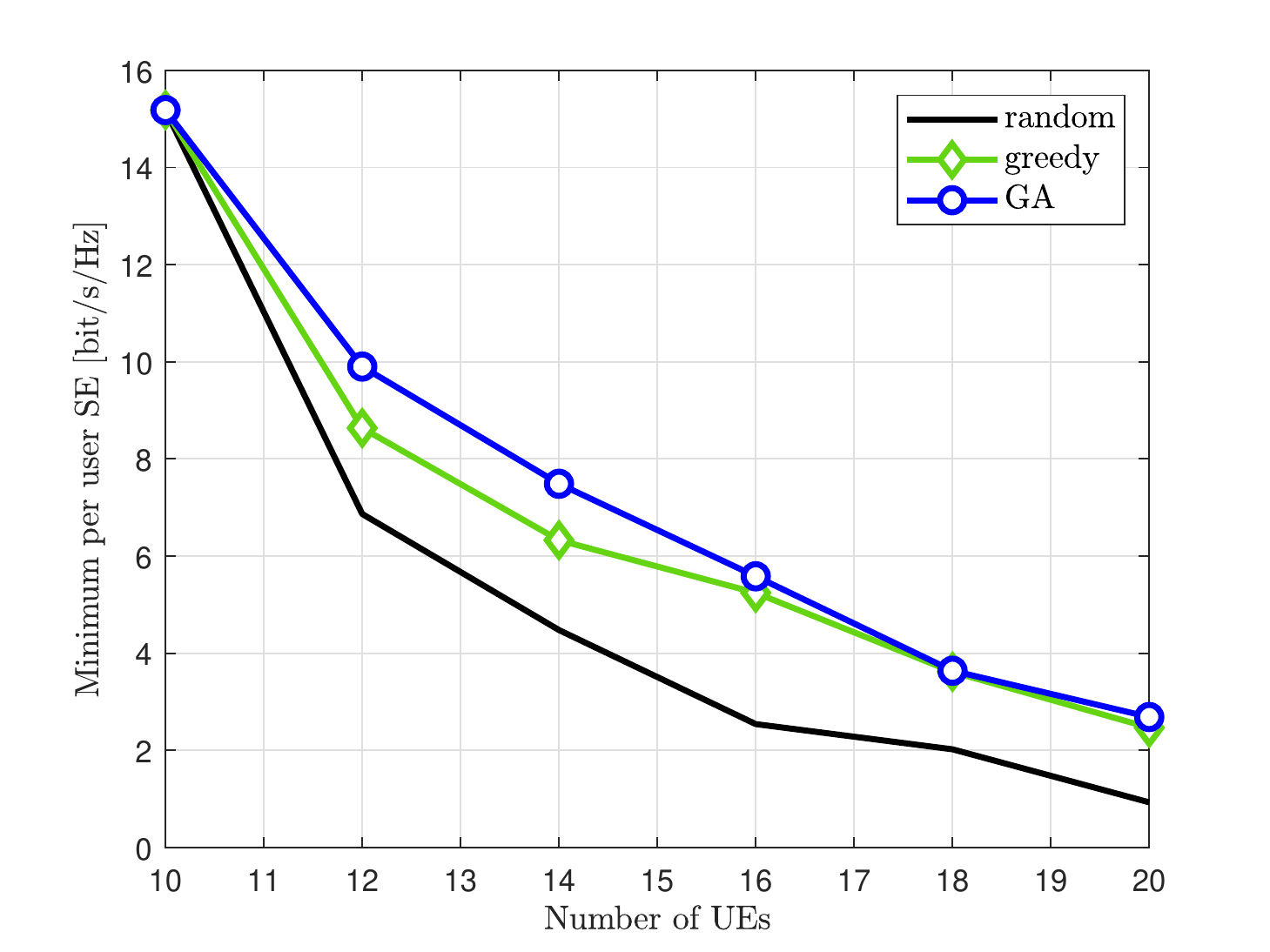}}}\hfill
\caption{a) Average sum SE, b) average per user SE and c) average minimum per user SE, as a function of the number of UEs. Parameter values: $\tau_\text{p}=10$, $L=50$ and $N=4$.} 
\label{fig:SExK}
\end{figure}

Fig. \ref{fig:SExK} presents the average values for sum SE, per user SE and minimum per user SE, varying the number of UEs, with $\tau_\text{p}=10$, $L=50$ and $N=4$. From Fig. \ref{fig:minSExK}, we confirm the effectiveness of the proposed GA scheme for improving the minimum SE among the $K$ UEs, compared to the greedy and mainly to the random PA strategies, while it did not improve the average per user SE and the average sum SE compared to the greedy PA, only compared to the random approach. Finally, we can observe that the GA-based PA is a suitable way to make the QoS more uniform in XL-MIMO systems and that the channel estimation NMSE is promising in terms of performance-complexity trade-off to evaluate each candidate on the population in each iteration of the Algorithm \ref{alg:PA:GA}. It reinforces the conjecture that one can make intuitively that the SINR of the active UEs can be improved considerably by reducing the channel estimate error as a consequence of lowering the pilot contamination due to an appropriate pilot assignment and subarray joint selection procedures.

\section{Conclusions}
\label{sec:Conclusions}
We have investigated the performance of different pilot assignment schemes to improve the QoS uniformity in XL-MIMO systems under both LoS and NLoS channel components. \colb{A proper PA strategy is paramount in crowded XL-MIMO systems, where the number of available orthogonal pilots is much smaller than the number of active UEs. Hence,} a realistic XL-MIMO spatially correlated channel model has been introduced, considering that the XL-MIMO subarray antennas are not co-located, meaning that a given UE may be close to some SAs and distant from others. Consequently, in most cases, only some of the SAs are relevant to the signal detection of a given UE, such that a \colb{simple} SA selection strategy \colb{based on Strongest-UE was implemented} to make the system scalable without compromising the SE. \colb{The proposed GA-based PA method for SA-based XL-MIMO systems was compared with three other methods: a) the genie PA, which is unfeasible in most practical scenarios of applications, as it is an exhaustive search for the best solution in terms of minimizing the channel estimation errors; b) the greedy PA, which is a low complexity algorithm and was extensively deployed in the literature; c) and the random PA, the simplest method, treated as the worst SE performance benchmark. The proposed GA-based PA strategy for XL-MIMO systems} considerably improved the minimum per-user spectral efficiency compared to the greedy and random PA methods, achieving nearly-optimal results under affordable computational complexity. \colb{This is in line with the requirements of beyond 5G networks, where making the quality of service more uniform is more important than improving the peak data rates. Also, the proposed PA strategy also provided higher sum rates than the random and greedy strategies.} Furthermore, the proposed way for evaluating each candidate in the GA population via the channel estimation NMSE analytical expression has performed remarkably well \colb{with respect to improving the SE, especially the minimum per-user SE.}

\appendices

\section{Proof of Eq. \eqref{h_NLoS_distribution} and \eqref{R}}
\label{app1}

{For notation simplicity, we will use $\textbf{a}_{kl}^{(b)}=\textbf{a}(\varphi_{kl}^{(b)},\theta_{kl}^{(b)})$ and $\textbf{a}_{kl}^{(0)}=\textbf{a}(\overline{\varphi}_{kl},\overline{\theta}_{kl})$. First of all, it is simple to find that $\textbf{h}_{kl}^\text{NLoS}$, as defined in \eqref{h_NLoS}, is a zero mean random vector, since $g_{kl}^{(b)}$ and $\textbf{a}_{kl}^{(b)}$ are independent and therefore uncorrelated:}
\begin{align*}
{
\mathbb{E}\{\textbf{h}_{kl}^\text{NLoS}\} =
\sum_{b=1}^B \mathbb{E}\{ g_{kl}^{(b)} \textbf{a}_{kl}^{(b)} \} = \sum_{b=1}^B \underbrace{\mathbb{E}\{ g_{kl}^{(b)} \}}_{0} \mathbb{E}\{ \textbf{a}_{kl}^{(b)} \} = 0.
}
\end{align*}

{Second, the covariance matrix of $\textbf{h}_{kl}^\text{NLoS}$ is computed as}
\begin{align}
{
\textbf{R}_{kl}
} { ~=~ } & { 
\mathbb{E}\{\textbf{h}_{kl}^\text{NLoS}(\textbf{h}_{kl}^\text{NLoS})^\text{H}\} - \mathbb{E}\{\textbf{h}_{kl}^\text{NLoS}\}\mathbb{E}\{(\textbf{h}_{kl}^\text{NLoS})^\text{H}\}
} \notag \\ { =~ } & { 
\mathbb{E}\{ \left[ \sum_{b=1}^B g_{kl}^{(b)} \textbf{a}_{kl}^{(b)} \right]\left[ \sum_{b=1}^B (g_{kl}^{(b)})^* (\textbf{a}_{kl}^{(b)})^\text{H} \right] \} - 0\cdot0
} \notag \\ { =~ } & { 
\sum_{b=1}^B \mathbb{E}\{|g_{kl}^{(b)}|^2\}\mathbb{E}\{\textbf{a}_{kl}^{(b)}(\textbf{a}_{kl}^{(b)})^\text{H}\}~+
} \notag \\ & {
\sum_{b=1}^B\sum\limits_{\substack{b'=1\\b'\neq b}}^B \underbrace{\mathbb{E}\{g_{kl}^{(b)}(g_{kl}^{(b')})^*\}}_{0}\mathbb{E}\{\textbf{a}_{kl}^{(b)}(\textbf{a}_{kl}^{(b')})^\text{H}\},
}
\label{h_NLoS_corr}
\end{align}
{where $\mathbb{E}\{g_{kl}^{(b)}(g_{kl}^{(b')})^*\}=0$ because $g_{kl}^{(b)}$, $b=1,\dots,B$, are i.i.d. random variables. Since the array response vector is given by $[\textbf{a}(\varphi,\theta)]_n=e^{-j2(n-1)\pi\frac{d}{\lambda}\frac{\text{sin}(\varphi)}{{\text{cos}(\theta)}}}$, the matrix $\textbf{a}_{kl}^{(b)}(\textbf{a}_{kl}^{(b)})^\text{H}$ can be calculated as}
\begin{equation*}
{
[\textbf{a}_{kl}^{(b)}(\textbf{a}_{kl}^{(b)})^\text{H}]_{m,n} = \text{exp}\left(j2(m-n)\pi\frac{d}{\lambda}\frac{\text{sin}(\varphi_{kl}^{(b)})}{\text{cos}(\theta_{kl}^{(b)})}\right),
}
\end{equation*}

{The expectation $\mathbb{E}\{\textbf{a}_{kl}^{(b)}(\textbf{a}_{kl}^{(b')})^\text{H}\}\}$, over the angles $\varphi_{kl}^{(b)}$ and $\theta_{kl}^{(b)}$, is the same for $b=1,\dots,B$, and by definition is}
\begin{align*}
{
[\mathbb{E}\{\textbf{a}_{kl}^{(b)}(\textbf{a}_{kl}^{(b)})^\text{H}\}]_{m,n} 
} & { = 
\int\int [\textbf{a}_{kl}^{(b)}(\textbf{a}_{kl}^{(b)})^\text{H}]_{m,n} f_{kl}(\varphi,\theta)d\varphi d\theta
} \\ & { =
\int\int e^{j2(m-n)\pi\frac{d}{\lambda}\frac{\text{sin}(\varphi)}{\text{cos}(\theta)}}f_{kl}(\varphi,\theta)d\varphi d\theta.
}
\end{align*}

Thus, this term can go out of the summary in \eqref{h_NLoS_corr}. Finally, recalling that $\beta_{kl}^\text{NLoS}=\sum_{b=1}^B \mathbb{E}\{|g_{kl}^{(b)}|^2\}$:
\begin{align*}
{
[\textbf{R}_{kl}]_{m,n} =}~& {
\beta_{kl}^\text{NLoS}
\int\int e^{j2(m-n)\pi\frac{d}{\lambda}\frac{\text{sin}(\varphi)}{\text{cos}(\theta)}}f_{kl}(\varphi,\theta)d\varphi d\theta.
}
\end{align*}
	
\bibliographystyle{IEEEtran}
\bibliography{ref}


\end{document}